\documentclass[
reprint,
nofootinbib,
 amsmath,amssymb,
 aps,
]{revtex4-2}
\usepackage[utf8]{inputenc}
\usepackage{listings}

\usepackage{graphicx} 
\graphicspath{{figures}}

\usepackage{tikz}
\usetikzlibrary{quantikz2}

\usepackage{xcolor}
\usepackage{graphicx}
\usepackage{dcolumn}
\usepackage{bm}
\usepackage{bbm}
\usepackage{amsmath}
\usepackage{orcidlink}
\usepackage{physics}
\usepackage{natbib}
\usepackage[version=4]{mhchem}
\usepackage{gensymb}
\makeatletter
\let\old@makecaption=\@makecaption
\usepackage{subcaption}
\let\@makecaption=\old@makecaption
\makeatother
\usepackage{siunitx}
\usepackage{graphicx,multirow}
\usepackage{array}
\usepackage{hyperref}
\hypersetup{
    colorlinks,
    linkcolor={red!80!black},
    citecolor={blue!80!black},
    urlcolor={blue!80!black}
}
\usepackage[capitalize]{cleveref}
\crefname{section}{Sec.}{Secs.}
\usepackage[super]{nth}
\usepackage{lipsum}

\usepackage{dsfont}
\newcommand\blfootnote[1]{%
  \begingroup
  \renewcommand\thefootnote{}\footnote{#1}%
  \addtocounter{footnote}{-1}%
  \endgroup
}
\begin{document}
\title{Classical and quantum cost of measurement strategies for quantum-enhanced auxiliary field Quantum Monte Carlo}
\date{\today}

\author{Matthew Kiser\orcidlink{0000-0002-9357-7583}$^{1,2,\dagger}$}
\affiliation{QUTAC Material Science Working Group}

\author{Anna Schroeder\orcidlink{0009-0006-4971-3265}$^{3,4,\dagger}$}
\affiliation{QUTAC Material Science Working Group}

\author{Gian-Luca R. Anselmetti\orcidlink{0000-0002-8073-3567}$^{5}$}
\affiliation{QUTAC Material Science Working Group}

\author{Chandan Kumar\orcidlink{0000-0001-6510-4204}$^{6}$}
\affiliation{QUTAC Material Science Working Group}

\author{Nikolaj Moll\orcidlink{0000-0001-5645-4667}$^{5}$}
\affiliation{QUTAC Material Science Working Group}

\author{Michael Streif\orcidlink{0000-0002-7509-4748}$^{5,\ast}$}
\affiliation{QUTAC Material Science Working Group}

\author{Davide Vodola\orcidlink{0000-0003-0880-3548}$^{7}$}
\affiliation{QUTAC Material Science Working Group}

\begin{abstract}
Quantum-enhanced auxiliary field quantum Monte Carlo (QC-AFQMC) uses output from a quantum computer to increase the accuracy of its classical counterpart. The algorithm requires the estimation of overlaps between walker states and a trial wavefunction prepared on the quantum computer. We study the applicability of this algorithm in terms of the number of measurements required from the quantum computer and the classical costs of post-processing those measurements. We compare the classical post-processing costs of state-of-the-art measurement schemes using classical shadows to determine the overlaps and argue that the overall post-processing cost stemming from overlap estimations scales like $\mathcal{O}(N^9)$ per walker throughout the algorithm. With further numerical simulations, we compare the variance behavior of the classical shadows when randomizing over different ensembles, e.g., Cliffords and (particle-number restricted) matchgates beyond their respective bounds, and uncover the existence of covariances between overlap estimations of the AFQMC walkers at different imaginary time steps. Moreover, we include analyses of how the error in the overlap estimation propagates into the AFQMC energy and discuss its scaling when increasing the system size. 
\end{abstract}
\maketitle



\section{Introduction}
\blfootnote{
$^\dagger$ Contributed equally; Alphabetical otherwise\\
$^\ast$ Corresponding author: michael.streif@boehringer-ingelheim.com\\
$^1$ Volkswagen AG, Wolfsburg, Germany\\
$^2$ TUM School of Natural Sciences, Technical University of Munich, Garching, Germany\\
$^3$ Merck KGaA, Frankfurter Stra{\ss}e 250, 64293 Darmstadt, Germany\\
$^4$ Quantum Computing Group, Department of Computer Science, Technical University of Darmstadt, Darmstadt, Germany\\
$^5$ Quantum Lab, Boehringer Ingelheim, Ingelheim am Rhein, Germany\\
$^6$ BMW Group, Garching, Germany\\
$^7$  BASF Digital Solutions GmbH, Next Generation Computing, Pfalzgrafenstr. 1, 67056, Ludwigshafen, Germany}
\\

More accurate and faster solutions to the electronic structure problem are needed across various industry-relevant use cases, such as designing new drugs, developing better cathode materials and finding more efficient synthetic pathways of fertilizers. Quantum computing is believed to provide an avenue for more accurate simulations of chemistry and materials science~\cite{Cao2019Quantum, McArdle2020Quantum, dalzell2023quantum}.

Quantum algorithms and hardware can be classified under two umbrella terms: noisy intermediate-scale quantum (NISQ) \cite{preskill2018quantum} and fault-tolerant quantum computing (FTQC). The latter indicates a true paradigm shift with rigorously proven quantum advantages and promising avenues for industrial applications \cite{santagati2023drug,bauer2020quantum}. In contrast, in the current NISQ era, proven quantum advantages for applications are not available. However, with the FTQC era still many years away, it is worth investigating the usefulness of NISQ algorithms for applications in industry. 
Due to present errors on NISQ-era computers, NISQ algorithms, such as the Variational Quantum Eigensolver (VQE) \cite{peruzzo2014variational}, are limited to shallow circuits acting on a small number of qubits. This restricts the application of those algorithms to small molecules with only tens of spin orbitals, in turn limiting the range of applicability of NISQ algorithms in industry.

A second major challenge of applying NISQ algorithms to industrial applications is the extraction of observables from the quantum computer through quantum measurements. As quantum measurements are probabilistic, numerous samples are required to estimate expectation values accurately. Recently, there have been many developments in improving the measurement bottleneck using, for example, randomized measurements~\cite{elben2023randomized} such as classical shadows~\cite{huang2020predicting, hadfield2020measurements, gresch2023guaranteed, zhao2020fermionic, wan2022matchgate, low2022classical}.

In the present work, we investigate a NISQ algorithm called quantum-enhanced auxiliary field quantum Monte Carlo (QC-AFQMC) \cite{huggins2022unbiasing}. This method aims to enhance the computational accuracy of classical auxiliary field quantum Monte Carlo (AFQMC)~\cite{blankenbecler1981monte, motta2018abinitio} by utilizing a quantum trial state wavefunction generated on a NISQ-era computer~\cite{huggins2022unbiasing}. A key aspect of this quantum algorithm is the estimation of overlaps between the walker in AFQMC, which are non-orthogonal Slater determinants, and the trial wavefunction. Different schemes for evaluating the overlaps have been recently investigating. For example, in Ref.~\cite{xu2023quantumassisted}, a general Bayesian inference method has been introduced to reduce the variance of the energy computed via AFQMC with finite number of measurements.

Here, we discuss the prevalent measurement challenge in NISQ algorithms in the context of estimating the overlaps in QC-AFQMC and investigate the scaling of the required measurements to obtain error bounds of the error in the final AFQMC energy estimates when increasing the system size. Additionally, we investigate the classical post-processing cost to use the measurement outcomes to determine the overlaps.

The paper is organized as follows. In \cref{sec:qc_afqmc}, we provide an overview of the QC-AFQMC algorithm.  In \cref{sec:trial_wfn}, we discuss how to prepare a trial wavefunction using the variational quantum eigensolver (VQE) on the quantum computer. In \cref{sec:overlaps_in_afqmc}, we highlight the steps of the algorithm where estimating overlaps between walker states and the trial wavefunction are required. In \cref{sec:measurement_methods}, we discuss the various techniques for estimating overlaps when using a quantum computer.  In \cref{sec:statistical_properties}, we discuss the statistical behavior associated with the overlap estimation using the different measurement schemes. In \cref{sec:qc_afqmc_finite}, we introduce an error model, which allows us to perform numerical simulations of QC-AFQMC under a finite number of measurements, and  estimate the ground state energy of small molecular systems. Finally, in \cref{sec:costs} we discuss the quantum and classical computational costs of the algorithm. 

\section{Quantum enhanced Auxiliary field quantum Monte Carlo (QC-AFQMC)}\label{sec:qc_afqmc}

In this section, we briefly introduce the Auxiliary-Field Quantum Monte Carlo (AFQMC) technique and its quantum counterpart QC-AFQMC and refer the reader to the literature~\cite{zhang2013auxiliary, motta2018abinitio} for more details.

We consider the electronic Hamiltonian in the second quantization
\begin{equation}
\begin{split}
{H} & = {H}_1 + {H}_2 \\
&=\sum_{pq} h_{pq} {a}_{p}^{\dagger} {a}_{q} + \frac{1}{2} \sum_{pqrs} v_{pqrs}{a}_{p}^{\dagger} {a}_r {a}^{\dagger}_{q} {a}_s \, ,
\end{split}
\end{equation}
where ${a}_{p}^{\dagger}$ and ${a}_{q}$ are fermionic creation and annihilation operators of orbitals $p$ and $q$, respectively. The indices $p,q,r,s$ run over the set $[N] = \{1,\dots, N\}$ where $N$ is the total number of spin orbitals. The terms $h_{pq} $ and $v_{pqrs}$ are the matrix elements of the one-body $H_1$ and two-body interactions ${H}_2$ of ${H}$, respectively. Here, we omit the spin indices for simplicity. 

AFQMC relies on decomposing the two-body interactions $H_2$ in terms of sum of squares of one-body operators ${v}_\gamma$ such that the Hamiltonian ${H}$ becomes
\begin{equation}\label{eqn_cholesky_hamilonian}
{H} = {v}_0 - \frac{1}{2}\sum_{\gamma=1}^{N_\gamma}   {v}_\gamma^2
\end{equation}
with ${v}_0 = {H}_1 $ and $
{v}_\gamma = i \sum_{pq} L^{\gamma}_{pq} {a}_{p}^{\dagger}{a}_{q}.
$
The $N_\gamma$ matrices $L^{\gamma}_{pq}$ are obtained via a Cholesky decomposition of the two-body matrix elements $v_{pqrs}$ via $v_{pqrs} = \sum_{\gamma=1}^{N_\gamma} L^{\gamma}_{pr} L^{\gamma}_{qs}$.

The ground state wavefunction of $H$ is obtained from the imaginary propagation of an initial state $\ket{\Psi_{I}}$ 
\begin{equation}\label{eqn_afqmc_imaginary_proj}
\ket{\Psi_0} = \lim_{n \to \infty} \frac{\left[ e^{-\Delta\tau \left(  H - E_0 \right) }  \right]^{n} \ket{\Psi_{I}}}{\matrixel{\Psi_{I}}{e^{-2 n \Delta\tau \left(  H - E_0 \right) } }{\Psi_{I}}}\, ,
\end{equation}
where $\Delta\tau$ is the imaginary time step, and $E_0$ is the unknown ground-state energy, which can be estimated adaptively as the calculation progresses. Here we assume that $\braket{\Psi_0}{\Psi_{I}} \neq 0$ and we choose $\ket{\Psi_{I}}$ to be a Slater determinant.

To find an efficient implementation of the exponential in \cref{eqn_afqmc_imaginary_proj}, a Hubbard-Stratonovich transformation~\cite{motta2018abinitio, negele2018quantum} is used
\begin{equation}
\label{eq:hstrafo}
    e^{-\frac{\Delta\tau}{2} \sum_\gamma {v}_{\gamma}^{2}} = \int \text{d}\mathbf{x} \; p(\mathbf{x}) \; e^{-i \sqrt{\Delta\tau} \sum_\gamma x_\gamma {v}_\gamma} + \mathcal{O}(\Delta\tau^2)~, 
\end{equation}
where $p(\mathbf{x})$ is the standard normal distribution and $\mathbf{x} \in \mathds{R}^{N_\gamma}$ are the auxiliary fields. 
The effect of this transformation is to map the actual interaction to an integral over one-body evolution operators. In this way, the state of the system at imaginary time step $n$ of the propagation can be described by an ensemble of $N_{w}$ Slater determinants $\{\ket*{\phi^{(n)}_z}, z=1,\dots,N_{w}; n=1,\dots, N_{\Delta\tau}\}$ ($N_{\Delta\tau}$ being the total number of imaginary time steps) with weights $\{W_{n,z}\}$ and phases 
$\{e^{i\theta_{n,z}}\}$. The ensemble of  Slater determinants is generally referred to as random walkers.

In the so-called free-projection AFQMC~\cite{motta2018abinitio}, the walker states, their weights, and phases are updated according to
\begin{gather}
\ket*{ \phi^{(n+1)}_z} = {B} ({\bf x}_{n,z}) \ket*{ \phi^{(n)}_z}  \\
W_{n+1,z} \, e^{i\theta_{n+1,z}} = \frac{ \braket*{ \Psi_\mathrm{T}}{\phi^{(n+1)}_z } }{ \braket*{ \Psi_\mathrm{T} }{ \phi^{(n)}_z } }\,W_{n,z} \, e^{i\theta_{n,z} } \, ,
\end{gather}
where the operator
\begin{equation}
{B}({\bf x}) = \exp[-\Delta\tau\,\left(-E_{0}+{H}_1\right)+ \sqrt{\Delta\tau}\sum_{\gamma}x_{\gamma}{v}_{\gamma}]
\end{equation}
contains only one-body terms. The overlaps $\braket*{ \Psi_\mathrm{T}}{\phi^{(n)}_z }$ are computed by using a trial state $\ket{\Psi_\mathrm{T}}$. Ref.~\cite{huggin2022unbiasing} has shown that quantum computers enable us to explore trial wavefunctions $\ket{\Psi_\mathrm{T}}$  beyond what is classically tractable, enhancing the performance of classical AFQMC calculations. With this approach, a 16-qubit experiment has been performed~\cite{huggins2022unbiasing}.

Different forms of the Hubbard-Stratonovich transformation exist, and they can improve the performance of AFQMC~\cite{motta2018abinitio}. One of them is called mean-field subtraction and introduces a shift ${v}_\gamma \to {v}_{\gamma}-\overline{v}_{\gamma}$ in the one body operators ${v}_\gamma$ in the Hamiltonian in \cref{eqn_cholesky_hamilonian} where the constant  $\overline{v}_{\gamma}$ are expectation values on the trial state
\begin{equation}\label{eqn_mean_field_subt}
\overline{v}_{\gamma} = \matrixel*{\Psi_\mathrm{T}}{{v}_{\gamma}}{\Psi_\mathrm{T}}
\end{equation}
This reduces fluctuations and minimizes systematic errors during the free-projection calculation~\cite{al-saidi2006_auxiliary, purwanto2005_correlation}.

During the AFQMC propagation, the walker ensemble allows us to access observables of the system. For example, the total energy is computed as
\begin{equation}\label{eqn_total_energy}
    \mathcal{E}^{(n)} = \frac{\sum_{z=1}^{N_{w}}W_{n,z}e^{i\theta_{n,z}}\mathcal E_{\text{loc}}(\phi^{(n)}_z)}{\sum_{z=1}^{N_{w}}W_{n,z}e^{i\theta_{n,z}}} \, ,
\end{equation}
where $\mathcal{E}_{\text{loc}}(\phi^{(n)}_z)$ is the local energy
\begin{equation}
    \mathcal{E}_{\text{loc}}(\phi^{(n)}_z) = \frac{\matrixel*{\Psi_\mathrm{T}}{{H}}{\phi^{(n)}_z}}{\braket*{\Psi_\mathrm{T}}{\phi^{(n)}_z}}\label{eqn_local_energy}
\end{equation}
defined as the mixed expectation value of the Hamiltonian with the trial state $\ket{\Psi_\mathrm{T}}$.

The energy estimator in \cref{eqn_total_energy} often experiences an exponential increase in its variance due to the fermionic sign problem~\cite{zhang2003quantum}. The phaseless approximation was introduced in Ref.~\cite{zhang2003quantum} to prevent this exponential growth. Following Ref.~\cite{malone2022ipie}, the projection is  accomplished by changing the updating rule of the walkers to 
\begin{gather}\label{eqn_phaseless_approx}
\ket*{ \phi^{(n+1)}_z} = {B} ({\bf x}_{n,z} -\overline{{\bf x}}_{n,z}) \ket*{ \phi^{(n)}_z}  \\
W_{n+1,z}= I_\text{ph}({\bf x}_{n,z},\overline{{\bf x}}_{n,z}; \phi^{(n+1)}_z)\, W_{n,z} 
\end{gather}
where the components of vector  $\overline{{\bf x}}_{n,z}$, called the force bias, are defined as 
\begin{equation}\label{eqn_force_bias}
(\overline{{\bf x}}_{n,z})_{\gamma}=-\sqrt{\Delta\tau}\frac{\matrixel*{\Psi_\mathrm{T} }{{v}_{\gamma}}{\phi^{(n)}_z}}{\braket*{\Psi_\mathrm{T} }{ \phi^{(n)}_z}}.
\end{equation}
The function $I_\text{ph}$ is defined as
\begin{equation}
\begin{split}\label{eqn:importance_ipie}
I_\text{ph}({\bf x}_{n,z},\overline{{\bf x}}_{n,z}; \phi^{(n+1)}_z) = |I (\mathbf{x}_{n,z},\mathbf{\bar{x}}_{n,z};\phi^{(n)})| \\ \times \
\text{max}(0, \cos(\theta_{n,z}))
\end{split}
\end{equation}
where $I$, called the importance function, is 
\begin{equation}
    I(\mathbf{x}_{n,z},\mathbf{\bar{x}}_{n,z};\phi^{(n)}) = O_{n,z}
    e^{\mathbf{x}_{n,z}\cdot\mathbf{\bar{x}}_{n,z}-\mathbf{\bar{x}}_{n,z}\cdot\mathbf{\bar{x}}_{n,z}/2},
 \label{eq:import}
\end{equation}
$O_{n,z}$ is the overlap ratio
\begin{equation}
O_{n,z} = \frac{\langle
    \Psi_\mathrm{T} |
    {B}(\mathbf{x}_{n,z}-\mathbf{\bar{x}}_{n,z}) | \phi^{(n)}_z
    \rangle}{
    \langle
    \Psi_\mathrm{T} | \phi^{(n)}_z
    \rangle},
 \label{eq:ovl}
\end{equation}
and the phase $\theta_{n,z}$ is given by
\begin{equation}\label{eqn_phase_diff_phaseless}
 \theta_{n,z} = \mathrm{arg}(O_{n,z}).
\end{equation}

\section{Generation of quantum trial wavefunctions}\label{sec:trial_wfn}

In this work, we use a variational quantum eigensolver (VQE)~\cite{peruzzo2014variational,mcclean2016theory,TILLY20221} to obtain approximate ground states; subsequently used as trial wavefunctions in AFQMC. Several VQE ans\"atze have been proposed to approximate ground states of molecular systems, such as the chemically inspired unitary coupled-cluster  ansatz~\cite{romero2018strategies}, hardware-efficient ans\"atze~\cite{Kandala2017Hardware-efficient} or the qubit coupled cluster ansatz~\cite{Ryabinkin2018qcc}. One major challenge in achieving accurate ground state energies is the high depth of those ans\"atze in the presence of errors on the quantum device \cite{Fedorov2022VQE_method}. Another challenge is ensuring  the correct particle number and spin symmetries in the ansatz. The quantum-number preserving (QNP) ansatz introduced in Ref.~\cite{Anselmetti_2021Local} provides a low-depth circuit which preserves the particle number, the number of alpha and beta electrons, as well as the total spin squared, ${S}^{2}$. In this work, we implement the QNP ansatz within the framework of PySCF \cite{PySCF:2018},  OpenFermion \cite{mcclean2019openfermion}, and cirq \cite{cirq_developers_2022_7465577}. To optimize the variational parameters, we use L-BFGS-B~\cite{liu1989limited} while employing finite differences to compute the first-order gradients. 

\section{Overlap calculations in QC-AFQMC\label{sec:overlaps_in_afqmc}}

\begin{figure}
    \centering
    \includegraphics[width=\columnwidth]{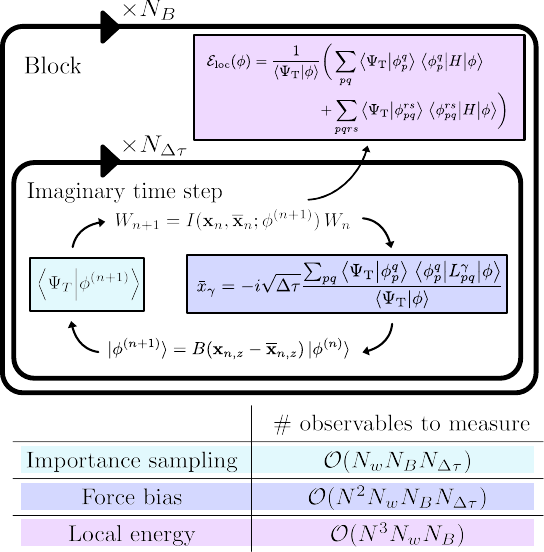}
    \caption{Visualization of the AFQMC algorithm. Colored terms are steps where we intend to measure overlaps between the trial wavefunction $\ket{\Psi_\mathrm{T}}$ and walker states $\ket{\phi}$ on a quantum device in QC-AFQMC. We note that the $\mathcal{O}(N^3)$ scaling in the local energy is achieved when exploiting Hamiltonian compression methods, such as double factorization or similar methods \cite{lee2021even,von2021quantum,huggins2021efficient}. Without these compression strategies, it would scale as $\mathcal{O}(N^4)$. }
    \label{fig:afqmcoverlaps}
\end{figure}

To utilize the output from a quantum computer as a trial wavefunction within AFQMC, overlap calculations between non-orthogonal Slater determinants $\ket*{\phi_z^{(n)}}$ and the state present on the quantum computer $\ket{\Psi_\mathrm{T}}$ must be performed at various stages of the algorithm, as highlighted in \cref{fig:afqmcoverlaps}. In the practical implementation of the algorithm \cite{malone2022ipie}, the evolution is split into $N_\mathrm{B}$-many \emph{blocks} where in each block the walkers are evolved by $N_{\Delta\tau}$ time steps $\Delta\tau$. Thus, the walkers are propagated $N_\mathrm{B}N_{\Delta\tau}$ times.

During the propagation of the walkers, \cref{eq:ovl}  must be evaluated for every walker to calculate its importance function. This requires $N_{w}N_\mathrm{B}N_\mathrm{\Delta\tau}$ overlap calculations throughout the AFQMC run.

Another instance of recurring overlap calculations is observed in the computation of the force bias, as defined in \cref{eqn_force_bias}. The matrix element of the $N_\gamma$ one-body Cholesky vectors between the trial wavefunction and the walker state must be calculated to determine the force bias. As highlighted in the Appendix of \cite{huggin2022unbiasing}, the force bias can be rewritten as
\begin{align}\label{eqn_force_bias_sd}
    \Bar{x}_\gamma=-i\sqrt{\Delta\tau}\frac{\sum_{pq}\braket{\Psi_\mathrm{T}}{\phi_{p}^{q}}{\matrixel{\phi_{p}^{q}}{L^{\gamma}_{pq}}{\phi}}}{\braket{\Psi_\mathrm{T}}{\phi}}\,,
\end{align}
where $\ket{\phi_{p}^{q}}$ denotes a singly excited walker state with an excitation from orbital $p$ to orbital $q$.  We omit here the indices $n$ and $z$ for the projection step and the walker, respectively. The calculation then requires the estimation of the overlaps $\braket{\Psi_\mathrm{T}}{\phi_{p}^{q}}$ on the quantum computer, while the Slater-Condon rule can be used to efficiently calculate $\matrixel{\phi_{p}^{q}}{L^{\gamma}_{pq}}{\phi}$ on the classical computer. 
Hence, the force bias calculation requires the quantum computation of $\mathcal{O}(N^2)$ overlaps for each walker, or $\mathcal{O}(N^2N_{w}N_\mathrm{B}N_\mathrm{\Delta\tau})$ overlaps in total.

A third instance of overlap calculation arises in the most important quantity of AFQMC, the local energy, as defined in \cref{eqn_local_energy}. Decomposing the local energy into its one- and two-body terms yields
\begin{align}\label{eqn_local_energy_sd}
    \frac{\matrixel{\Psi_\mathrm{T}}{H}{\phi}}{\braket{\Psi_\mathrm{T}}{\phi}}= \frac{1}{\braket{\Psi_\mathrm{T}}{\phi}}\biggl(&\sum_{pq}\braket{\Psi_\mathrm{T}}{\phi_p^q}\matrixel{\phi_p^q}{H}{\phi} \nonumber\\+ &\sum_{pqrs}\braket{\Psi_\mathrm{T}}{\phi_{pq}^{rs}}\matrixel{\phi_{pq}^{rs}}{H}{\phi}\biggr)\,,
\end{align}
where $\ket{\phi_p^q}$ and $\ket{\phi_{pq}^{rs}}$ denote singly and doubly excited walker states respectively. Again, the overlaps $\braket{\Psi_\mathrm{T}}{\phi_p^q}$ and $\braket{\Psi_\mathrm{T}}{\phi_{pq}^{rs}}$ can be calculated using the quantum computer. In contrast, the matrix elements $\matrixel{\phi_{pq}^{rs}}{H}{\phi}$ and $\matrixel{\phi_{p}^{q}}{H}{\phi}$ can be efficiently calculated on a classical computer using the Slater-Condon rule and the fact that the Hamiltonian $H$ is a two-body operator.
Without considering any exploration or compression techniques, the two-body part of the Hamiltonian consists of $\mathcal{O}(N^4)$ terms, which in turn requires an equal number of overlap calculations. Using factorization techniques, such as double factorization or similar methods \cite{lee2021even,von2021quantum,huggins2021efficient,motta2019efficient,lee2020stochastic}, reduces the scaling to $\mathcal{O}(N^3)$. In contrast to the overlap ratio and force bias, the local energy is calculated only periodically. Although this reduces the number of overlap calculations, it remains the most computationally demanding quantity for larger systems due to its $\mathcal{O}(N^3)$ scaling. In total, the calculation of the local energy would require $\mathcal{O}(N_{w}N^3N_\mathrm{B})$ overlap calculations in the entire algorithm.

We want to point out that determining the mean field shift, \cref{eqn_mean_field_subt}, also requires access to the trial wavefunction. Consequently, this quantity must be computed using the quantum state, too. However, unlike other quantities, this calculation cannot be efficiently translated to overlap calculations. Furthermore, it is worth mentioning that this quantity only needs to be computed once throughout the entire AFQMC run. In cases where the quantum state is generated via VQE, this quantity has already been calculated. Otherwise, if the state was generated differently and one would like to avoid measuring this expectation value, one could use a classical approximation to the trial wavefunction.

In the subsequent sections, we will discuss various approaches for computing the necessary overlaps using a quantum computer.

\section{Measurement methods}\label{sec:measurement_methods}
This section reviews the quantum measurement schemes employed in this work to obtain estimates of the overlap terms $\braket{\Psi_\mathrm{T}}{\phi_z}$, including the Hadamard test in \cref{subsec:Hadamard} and the classical shadows algorithm in \cref{subsec:classical_shadows_intro},  especially its variants: $N$-qubit Clifford shadows in \cref{subsubsec:cliff_shadows}, matchgate shadows in \cref{subsubsec:matchgate_shadows}, and orbital rotation shadows in \cref{subsubsec:low_shadows}. Readers familiar with the topic can proceed to the next section. The main takeaway is that different measurement methods exhibit varying guaranteed upper bounds to sampling complexity and post-processing costs when obtaining the desired quantities from the quantum device. In \cref{sec:statistical_properties}, we then report statistics of overlap estimates obtained from numerical simulations, while in \cref{sec:costs}, we focus on examining the concrete scaling behavior of these methods.

\subsection{Hadamard Test}\label{subsec:Hadamard}

\begin{figure}
\centering
\begin{quantikz}
\lstick{$\ket{0}$} & \gate{H} & \gate[style={dashed}]{S} & \ctrl{1} &\octrl{1} & \gate{H} &\meter{}\\
\lstick{$\ket{0^{\otimes N}}$} &\ghost{X} && \gate{U_{\Psi_\mathrm{T}}}  & \gate{U_{\phi_z}} & & 
\end{quantikz}
\caption{The quantum circuit representing the modified Hadamard test to calculate the overlap $\braket{\Psi_\mathrm{T}}{\phi_z}$ between the trial wavefunction $\ket{\Psi_\mathrm{T}}$ and a walker state $\ket{\phi_z}$ by using an ancilla qubit. The gate $H$ is a Hadamard gate, the unitaries $U_{\Psi_\mathrm{T}}$ and $U_{\phi_z}$ prepare the states $\ket{\Psi_\mathrm{T}}$ and $\ket{\phi_z}$, respectively. Depending on the application of the phase gate $S$, the real or imaginary part of the overlap is calculated.}
\label{fig:hadamard_circuit}
\end{figure}
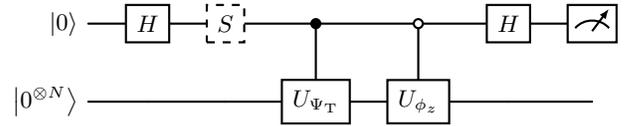

The Hadamard test \cite{cleve1998quantum,aharonov2005a,kitaev1995quantum} is a quantum algorithm that allows 
estimating expectation values $\expval{U}{\Psi} $ directly from a quantum processor with sampling complexity $\mathcal{O}(1/\epsilon^2)$ where $\epsilon$ is the additive error. The underlying idea is to encode the numerical value of $\expval{U}{\Psi} $ directly into the probability amplitude of an auxiliary qubit.
To implement the Hadamard test, one initializes an auxiliary qubit in $\ket{+}$ and applies a controlled $U$ gate between the auxiliary qubit and the target state $\ket{\Psi}$. When measuring the auxiliary qubit in the $X$ basis, we obtain the real part $\text{Re}(\expval{U}{\Psi})$ as the expectation value by associating outcomes $\ket{0}$ and $\ket{1}$ as $+1$ and $-1$, respectively. The imaginary part $\text{Im}(\expval{U}{\Psi})$ can be estimated with the same procedure, but the auxiliary qubit being initialized to $1/\sqrt{2}(\ket{0}-i\ket{1})$.

This approach can be adapted to estimate overlaps between two quantum states \cite{luongo2023chapter}, see \cref{fig:hadamard_circuit}, allowing its application in QC-AFQMC where we are interested in the overlap $\braket{\Psi_\mathrm{T}}{\phi_z}$. However, the Hadamard test necessitates access to controlled versions of state preparation unitaries $\ket{\Psi_\mathrm{T}} = U_{\Psi_\mathrm{T}} \ket{0}$ and $\ket{\phi_z} = U_{\phi_z} \ket{0}$ resulting in high circuit depths. Due to this condition, the Hadamard test is impractical in the current NISQ era. More resource-conscious modifications of the Hadamard test are studied in  \cite{huggins2020non,lu2021algorithms,russo2021evaluating}. 

Although the sampling complexity of the Hadamard test does not increase with the system size, it will still demand significant computing time on the quantum device as the circuits need to be executed for each walker at each time step. 
\subsection{Classical Shadows}\label{subsec:classical_shadows_intro}

Classical shadows, a scheme introduced in~\cite{huang2020predicting} based on the shadow tomography proposal \cite{aaronson2017shadow}, can predict many properties of a quantum system from very few measurements. Since its publication, various flavors \cite{zhao2020fermionic,low2022classical, wan2022matchgate} and applications to hybrid quantum-classical algorithms have been developed \cite{chan2022algorithmic,hadfield2020measurements,boyd2022training}. 
QC-AFQMC provides a natural application of the classical shadow framework when one has only limited quantum resources because the algorithm requires the calculation of many observables of the same state in the form of overlaps of the quantum trial wavefunction and classically computed walker states $\braket{\Psi_\mathrm{T}}{{\phi_{z}}}$. In the following, we summarize the general workflow of the classical shadow method as introduced in \cite{huang2020predicting}.
\begin{itemize}
    \item \textbf{Step 1:} Generate $n_{\mathrm{s}}$ copies of the $N$-qubit quantum state $\rho$, of which we want to estimate $M$ properties or observables $\{O_{i}\}_{i\in[M]}$.
    \item  \textbf{Step 2:} Measure each copy in a randomly rotated basis. In practice, this is equivalent to applying a randomly sampled unitary from a tomographically complete ensemble $U\sim \mathcal{U}$ and then measuring in the computational basis to gather $n_s$ bit strings $|\hat{b}\rangle \in \{0,1\}^N$. 
    \item \textbf{Step 3:} Generate \textit{snapshots} and eventually the \textit{classical shadows} from the measurement statistics by uncomputing $|\hat{b}\rangle$. When the ensemble $\mathcal{U}$ is chosen carefully, the mean estimate of the measurements is a quantum channel
    \begin{align}
        \mathcal{M} & (\rho) = \nonumber\\ &\mathcal{E}_{U \sim \mathcal{U}} \sum_{d \in \{0,1\}^N}
        \bra{d} U\rho U^\dagger \ket{d} U^\dagger \ketbra{d}{d} U~.
    \end{align}
    For each measurement outcome $|\hat{b}\rangle$, one obtains an unbiased estimator, the classical snapshot $\hat \rho$ of the quantum state $\rho$ by inverting this channel
    \begin{equation}
        \hat{\rho} = \mathcal{M}^{-1} \left(U^\dagger |\hat{b}\rangle\langle\hat{b}| U \right).
    \end{equation}
    The specific form of inverse channel $\mathcal{M}^{-1}$ is dependent on the choice of the distribution $\mathcal{U}$. The collection of all snapshots forms the \textit{classical~shadows}.
    \item \textbf{Step 4:} Predict the desired property $\expval{O_i}$ from the classical shadows. Generally, it is advised to use the median-of-means method to confine the estimator's variance \cite{huang2020predicting}. Thus, one divides the set of shadows into $K$ equally-sized parts and compute the $\rho$-estimator per subset $\hat{\rho}_{(k)}$ for $k\in [K]$. Then, the prediction of $\expval{O_i} = \Tr(O_i\rho)$ is the median of the set $\Tr(O_i\hat{\rho}_{(k)})_{k \in [K]}$.
\end{itemize}
Note that the last step is not necessary if the statistics of the estimator follow a Gaussian distribution, as is the case in this work (see \cref{fig:clifford_scaling_varince}). 
Ref.~\cite{huang2020predicting} provides rigorous statistical bounds to the shadow estimates. The variance of general shadows scales as
\begin{equation}
\mathrm{Var} [\hat{O}] = \mathcal{E} [(\hat{O} - \mathcal{E}[\hat{O}])^2] \leq
            \norm{O - \frac{\Tr(O)}{2^N}}^2_{\mathrm{shadow}} \, , 
\label{eq:variance_shadows}
\end{equation}
where the \textit{shadow norm} is defined as 
\begin{align}
    &\norm{O}_{\mathrm{shadow}} = \\ 
   & \max_{\sigma: \mathrm{state}} \sqrt{
        \mathcal{E}_{U \sim \mathcal{U}} \sum_{d \in \{0,1\}^n}
        \bra{d} U\sigma U^\dagger \ket{d} \bra{d} U \mathcal{M}^{-1}(O)U^\dagger 
        \ket{d}^2
    } \, . \nonumber
\end{align}

\subsubsection{$N$-qubit Clifford Shadows}\label{subsubsec:cliff_shadows}

 If we choose the ensemble $\mathcal{U}$ to be the set of all $N$-qubit Clifford operations, the inverse of the measurement channel becomes
    \begin{equation}
        \mathcal{M}^{-1} (U^\dagger |\hat{b}\rangle\langle\hat{b}| U) = \hat{\rho} 
        =  (2^N + 1) U^\dagger |\hat{b}\rangle\langle\hat{b}| U - \mathbbm{I}, 
    \end{equation} 
    and thus, the variance can be bounded with
    \begin{equation}
    \label{eq:clifford_var}
        \norm{O - \frac{\Tr(O)}{2^N}}^2_{\mathrm{shadow}} \leq 3\Tr(O^2) .
    \end{equation}
A numerical scheme to sample efficiently from $\mathcal{U}$ was devised in \cite{bravyi2020hadamard}, and made available through Qiskit \cite{Qiskit}.
The overlap estimation can be formulated in terms of Hermitian observables as illustrated in Ref.~\cite{huggins2022unbiasing}. This is achieved by incorporating a state orthogonal to the two states in question. Instead of the direct VQE output, one constructs
\begin{equation}\label{eq:tau_state}
    \ket{\tau} = \frac{\ket{0}+\ket{\Psi_\mathrm{T}}}{\sqrt{2}}
\end{equation} on the quantum computer. Then, using two projectors
    \begin{align}
        P_+ & = \ketbra{0}{\phi} + \ketbra{\phi}{0} \\
        P_- & = -i ( \ketbra{0}{\phi} - \ketbra{\phi}{0}) , 
    \end{align}
the overlap is estimated as 
\begin{align}
\label{eq:overlap}
      \braket{\phi}{\Psi_\mathrm{T}} & = \Tr \left[(P_+ + iP_-)\ketbra{\tau}{\tau}\right] \\ \nonumber
      & \approx 2(2^N + 1) \bra{\phi} \mathcal{E}_{U \sim \mathcal{U}}[U^\dagger |\hat{b}\rangle\langle\hat{b}| U\ket{0}]  .    
\end{align}
The upper bound of the variance of the estimate in the Clifford case is given by \cref{eq:clifford_var}, giving rise to a highly efficient sampling complexity of $\mathcal{O}(\log(M)/\epsilon^2)$ for $O = P_+ + iP_-$, where $M$ is the number of observables and $\epsilon$ is the desired additive error. Hence, in the context of QC-AFQMC, overlaps of all walkers at all time steps can be estimated from a small set of quantum measurements of $\ket{\tau}$. However, evaluating  \cref{eq:overlap} gives rise to an exponentially costly classical post-processing step due to the different bases of the Clifford shadows and the walker states, which are non-orthogonal Slater determinants. This issue was later addressed by Ref.~\cite{wan2022matchgate}, which proposes constructing shadows using matchgate tensors.

\subsubsection{Matchgate Shadows}\label{subsubsec:matchgate_shadows}

To circumvent the exponential post-processing step of Clifford shadows, a different form of shadows, known as matchgate shadows, was proposed~\cite{wan2022matchgate}. 

Instead of sampling from Clifford circuits, the distribution used for the shadows is \emph{matchgate circuits}. Matchgate circuits are described by \emph{fermionic Gaussian unitaries} (FGU) $U(Q)$ determined by $2N\times2N$ matrices from the real orthogonal group $Q\in O(2N)$, such that
\begin{equation}
    \label{eq:majorana_basischange}
    U(Q)^\dagger\gamma_\mu U(Q)=\sum_{\nu=1}^{2N}Q_{\mu\nu}\gamma_\nu~,
\end{equation}
where $\gamma_\mu$ are Majorana operators, $\gamma_{2j-1}=a_j+a_j^\dagger$ and $\gamma_{2j}=-i(a_j-a_j^\dagger)$. Matchgate circuits represent FGUs in quantum gates after the Jordan-Wigner transformation, which do not in general preserve the number of fermions $\eta$. This method efficiently calculates overlaps between a pure state, such as the trial wavefunction $\ket{\Psi_\mathrm{T}}$, and arbitrary Slater determinants, such as the walker states in QC-AFQMC. 

Similar to the Clifford circuit case, to calculate overlaps with Slater determinants, one prepares the state $\ket{\tau}$, see \cref{eq:tau_state}, on the quantum device.

To estimate the overlap between the trial wavefunction and a walker, one first calculates the coefficients $c_\ell$ in front of the monomial $z^\ell$ in the polynomial
\begin{equation}\label{eq:matchgateinversechannel}
    q(z)=\alpha_{\eta, N}\text{Pf}\left[\left.\left(C_{\boldsymbol{0}}+zW^*\tilde Q Q^\text{T}C_{\textbf{b}}Q\tilde Q^\text{T}W^\dagger\right)\right|_{\overline{S}_{\eta}}\right]~,
\end{equation}
where $\alpha_{\eta, N}=i^{\eta/2}/2^{N-\eta/2}$, Pf$(A)$ is the Pfaffian of a real-valued and  anti-symmetric matrix $A=-A^\text{T}$, $C_\textbf{i}$ is the covariance matrix of the computational basis state $\textbf{i}$, $W$ is a basis change from the set $\{\sqrt{2}a_1^\dagger,\sqrt{2}a_1,\dots,\sqrt{2}a_\eta^\dagger,\sqrt{2}a_\eta,\gamma_{2\eta+1},\dots,\gamma_{2N}\}$ to the set of Majorana operators $\{\gamma_\mu\}_{\mu\in[2N]}$ \cite{wan2022matchgate}, $\tilde Q$ encodes the Slater determinant, $Q$ is the matchgate rotation for a given snapshot and ${\overline{S}_{\eta}}=[2N]\setminus \{1,3,\dots,2\eta-1\}$. The Pfaffian is a polynomial with degree of at most $N-\eta/2$; thus, calculating the coefficients of $q(z)$ requires $\mathcal{O}((N-\eta/2)^4)$ time on a classical computer using polynomial interpolation \cite{wan2022matchgate}. The coefficients are then used to calculate the unbiased estimate $o^{(j)}$  ($j \in [n_s]$ with $n_s$ number of shadows) of the overlap with a Slater determinant using
\begin{equation}
    \label{eq:kianna_estimation}
    o^{(j)}=2\sum_{\ell=0}^N\binom{2N}{2\ell}\binom{N}{\ell}^{-1}c_\ell~.
\end{equation}
The final estimate of the overlap $\braket{\phi}{\Psi_\mathrm{T}}$ is given by averaging the unbiased estimators $\braket{\phi}{\Psi_\mathrm{T}} \approx \sum_j o^{(j)} / n_s$ or by median-of-means estimate~\cite{wan2022matchgate}.


The variance of this estimation is bounded by \cref{eq:kiannabound} and scales with $\mathcal{O}(\sqrt{N}\log N)$, which means that the number of samples required scales polynomially with system size. Thus, this method is both efficient in the number of samples and in its classical post-processing step \cite{wan2022matchgate}. 

Either the starting state $\ket{\Psi_\mathrm{T}}$ or the observables being measured must be even, in that they are in the span of the product space of operators with an even number of Majorana operators. These hold in most practical cases. And in the case of QC-AFQMC, if this condition is not fulfilled, one can transform the problem to satisfy it, as shown in Appendix A of \cite{wan2022matchgate}.

\subsubsection{Orbital rotation shadows}\label{subsubsec:low_shadows}

Shadow estimation schemes \cite{zhao2020fermionic,low2022classical} twirling over the particle number restricted ensemble of matchgates have been proposed. The latest \cite{low2022classical} recovers efficient estimation independent of $N$, only scaling in the number of occupied fermionic modes $\eta$, also referred to as the particle number of the trial state. When restricting FGUs to preserve particle number symmetry, one obtains the ensemble of single-particle basis rotations or orbital rotations $U(u)$ which are defined by restricting the  following transformations of the fermionic creation and annihilation operators
\begin{equation}
 U^\dagger(u) a_p^{\dagger} U(u)=\sum_{q=1}^{N}u_{pq}a_q^{\dagger}~,
\end{equation}
where $u \in \mathbb{C}^{N \times N}$ is a unitary matrix defining the change of basis of the fermionic modes, with their representation $U(u) \in \mathbb{C}^{2^N \times 2^N}$ in Hilbert space. We refer to the submatrix restricted to the $k$-particle subspace as $U_k \in \mathbb{C}^{\binom{N}{k}\times \binom{N}{k}}$.



In this notation, it is convenient to use expectation values of $k$-particle reduced density matrices ($k$-RDM) defined by 
\begin{equation}
\left\langle D_{\vec{q}}^{\vec{p}}\right\rangle=\operatorname{Tr}\left[D_{\vec{q}}^{\vec{p}} \rho\right], \quad D_{\vec{q}}^{\vec{p}} = \prod_i^{N\rightarrow} p_i a_{i}^{\dagger} \prod_j^{\leftarrow N} q_j a_{j} ~.
\end{equation}
for any quantum state $\rho$, where $\vec{p}$ ($\vec{q}$) are bitstring representations of the $p$,$q$,($r$,$s$) indices from Sec. \ref{sec:qc_afqmc} for general $k$-body excitations in normal ordering and index the column (row) of the $k$-RDM. More explicitly, $\vec{p}, \vec{q} \in \mathcal{S}_{N, k}$, where

\begin{equation}
\mathcal{S}_{N, k} = \left\{ \{p_1 \ldots p_i \ldots p_N\}:\forall p_i \in \{0,1\} , \sum_{i=1}^N p_i=k\right\} ~. 
\end{equation}
$p_i$ ($q_i)$ refer to the bit of $i$-th index of the bitstring $\vec{p}$($\vec{q}$).
As the evaluation of general $k$-RDMs is possible from these measurements we describe the following procedure for general $k$, the basis state overlaps can be calculated from the special case of the $(k=\eta)$-RDM as discussed below.

The single-shot estimator for the expectation value of a given $k$-RDM for a measurement string $b$ in the computational basis under the basis transformation generated by $u$ is given by \cite{low2022classical}
\begin{equation}
\label{eq:lowsingleshotestimator}
\left\langle\hat{D}_{\vec{q}}^{\vec{p}}\right\rangle=\left\langle\vec{p}\left|U_k\left(v_b u\right) E_{\eta, k} U_k^{\dagger}\left(v_b u\right)\right| \vec{q}\right\rangle \, ,
\end{equation}
where the basis states $ \bra{\vec{p}}$ and $\ket{\vec{q}}$ are defined by $\ket{\vec{p}} =  \bra{0} \prod_i^{\rightarrow N} p_i a_{i}^{\dagger}$ and $\ket{\vec{q}} =  \prod_i^{N\leftarrow} q_i a_{i}^{\dagger} \ket{0} $. The matrix $v_b$ implements a rotation of $b$ into the ordered string $\vec{r}^o =  \{11\cdots 00\} \:= \{1^{\otimes \eta}\}+ \{0^{\otimes (N-\eta)}\} \in \mathcal{S}_{N, \eta}$ and the operator $E_{\eta, k}$ is defined as
\begin{equation}
 E_{\eta, k} = \sum_{\vec{r} \in \mathcal{S}_{N, k}}|\vec{r}\rangle\langle \vec{r}| \frac{\left(\begin{array}{c}
\eta-s^{\prime} \\
k-s^{\prime}
\end{array}\right)\left(\begin{array}{c}
N-\eta+s^{\prime} \\
s^{\prime}
\end{array}\right)}{(-1)^{k+s^{\prime}}\left(\begin{array}{c}
k \\
s^{\prime}
\end{array}\right)} .
\end{equation}
with $s^\prime = \sum_i r^o_i r_i$ and $\vec{r}$ indexing the basis states in the $k$-particle subspace in the same notation as $\vec{p}$ and $\vec{q}$.

To estimate overlap with Slater determinants, one needs an additional register of ancillas equal to the particle number $\eta$ of $\ket{\Psi_\mathrm{T}}$ and must prepare a modified state $\left|\tau^\prime\right\rangle$ \cite{low2022classical}:
\begin{equation}
\left|\tau^\prime\right\rangle=\frac{\left|0^{\otimes N}\right\rangle_\text{S}\left|1^{\otimes \eta}\right\rangle_\text{A}+ |\Psi_\mathrm{T}\rangle_\text{S} |0^{\otimes\eta}\rangle_\text{A}}{\sqrt{2}},
\end{equation}
 where $|\rangle_\text{S}$ refers to the original register and $|\rangle_\text{A}$ to the added ancilla register. This is due to the state $|\Psi_\mathrm{T}\rangle$ fully occupying the subspace of $\eta$ particles in general, therefore one has to add additional 'space' for a reference state as the ancilla register as the reference state used in Eq. \ref{eq:tau_state} has the wrong number of particles.


The overlaps can then be inferred from the $(k=\eta)$-RDM by~\cite{low2022classical}
\begin{equation}
\langle\vec{q} \, | \Psi_\mathrm{T}\rangle = 2\operatorname{Tr}\left[D_{\vec{q'} }^{\vec{p'} }\left|\tau^{\prime}\right\rangle\left\langle\tau^{\prime}\right|\right] ,
\label{eq:overlapfromrdmlow}
\end{equation}
with $\vec{p'} = \{0^{\otimes N}\}+\{1^{\otimes \eta}\}$ and $\vec{q'} = \vec{q}+ \{0^{\otimes \eta}\}$ and $\vec{p'},\vec{q'} \in  \mathcal{S}_{N+\eta, \eta}$.
The average shot variance of the estimator is then bounded by
\begin{equation}
\label{eq:lowbound}
    \mathbb{E}_{\vec{p}, \vec{q}}\left[\operatorname{Var}\left[\left\langle\hat{D}_{\vec{q}}^{\vec{p}}\right\rangle\right]\right] \leq\left(\begin{array}{l}
\eta \\
k
\end{array}\right)\left(1-\frac{\eta-k}{N}\right)^k\left(\frac{1+N}{1+N-k}\right),
\end{equation}
which scales as $\mathcal{O}(\eta^2)$. The estimator in \cref{eq:lowsingleshotestimator} is only classically efficient for constant $k$. For efficient estimation in $k$ one can adapt a scheme based on the polynomial interpolation of Pfaffians introduced in \cite{wan2022matchgate}.

\section{Statistical properties of overlap estimations}\label{sec:statistical_properties}

In the following, we discuss the statistical properties of the measurement methods reviewed in the previous section. Specifically, we focus on their application in estimating overlaps between non-orthogonal Slater determinants and a trial wavefunction.
\subsection{Hadamard test}

The statistical properties of the Hadamard test are well-established~\cite{polla2023optimizing}. When employed to compute the real  component of the overlap, the variance of the estimate is given by
\begin{align}
    \mathrm{Var}[\mathrm{Re}(\braket{\Psi_\mathrm{T}}{\phi})] = \frac{1-|\mathrm{Re}(\braket{\Psi_\mathrm{T}}{\phi})|^2}{n_m}\,,
\end{align}
where  $n_m$ denotes the number of circuit repetitions for a single overlap estimation. An analogous equation holds for the imaginary part. 
It is important to note that since we are dealing with a continuous set of walker states $\{\ket*{\phi^{(n)}_z}\}$, the Hadamard test needs to be repeated for each overlap calculation, and the measurements cannot be reused.
By utilizing the amplitude estimation algorithm, the Hadamard test could achieve Heisenberg scaling \cite{brassard2002quantum, lin2022lecture}; however, this comes at the expense of longer quantum circuits.

\subsection{Classical Shadows}\label{subsec:classical_shadows}

Using classical shadows to calculate the overlaps presents a significant advantage over the Hadamard test, as it allows for collecting the measurement outcomes prior to initiating the AFQMC calculations. This approach not only minimizes the idle time required to compute new overlaps between imaginary time steps but also reduces the number of measurements required. All measurements can be used to calculate  the overlaps between the trial wavefunction and all walker states, rather than just a single overlap with a walker state.

\begin{figure}
    \centering  \includegraphics[width=1\linewidth]{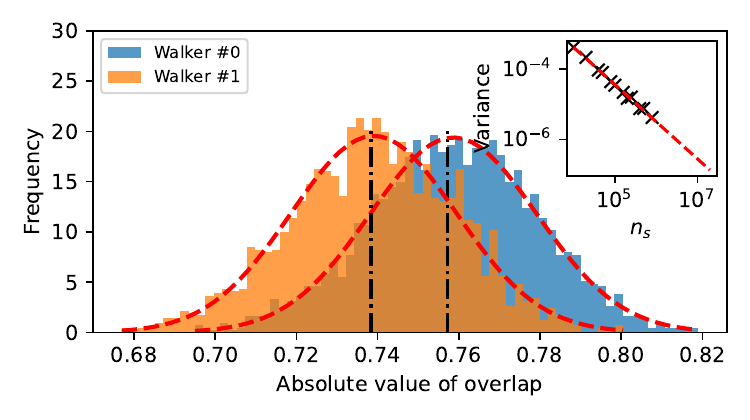}    \caption{Estimating overlaps with matchgate shadows. (main plot) The histograms show the distribution of the absolute values of the overlap estimates of two walker states w.r.t. a VQE output state ($\ce{H4}$) by conducting 2000 repeated shadow experiments, with each data point utilizing $10^4$ matchgate shadows. The dashed red lines represent two Gaussian distributions fitted to both distributions. (inset) The variance in the overlap distribution as the number of shadows increases.}
    \label{fig:matchgate_scaling_variance}
\end{figure}

To examine how the error in estimating overlaps $\braket{\Psi_\mathrm{T}}{\phi}$ scales with an increasing number of classical shadows, we utilize results from a simulation of the $\ce{H4}$ square with a side length of $1.23$\AA{} in STO-3G. Upon generating an approximate ground state using the VQE ansatz from Sec.~\ref{sec:trial_wfn}, we construct classical shadows using the $N$-qubit Clifford ensemble, matchgate shadows, and orbital rotation shadows. In 
\cref{fig:matchgate_scaling_variance} (main plot), we present the statistical analysis of estimating the overlap between the VQE output state and two arbitrarily chosen walker states from an AFQMC run using matchgate shadows. Each data point in the histogram signifies a single estimation of the overlap, employing $10^4$ shadows. We calculated the overlap $2000$ times to obtain statistics, with each calculation involving the sampling of new shadows. A fit of the data to a Gaussian distribution highlights that the error in the overlap estimations is normally distributed  \cite{scheurer23:_tailor_exter_correc_coupl_clust_quant_input}. In  the inset of \cref{fig:matchgate_scaling_variance} we investigate the scaling of the variance of the overlap estimation, \cref{eq:variance_shadows},  for one walker state as the number of snapshots increase. We observe the expected shot-noise scaling of classical shadows $n_{\mathrm{s}}^{-1/2}$ \cite{huang2020predicting}. In \cref{fig:clifford_scaling_varince}, we show the same finding for the case of Clifford shadows.

\begin{figure}
    \centering
    \includegraphics[width=1\columnwidth]{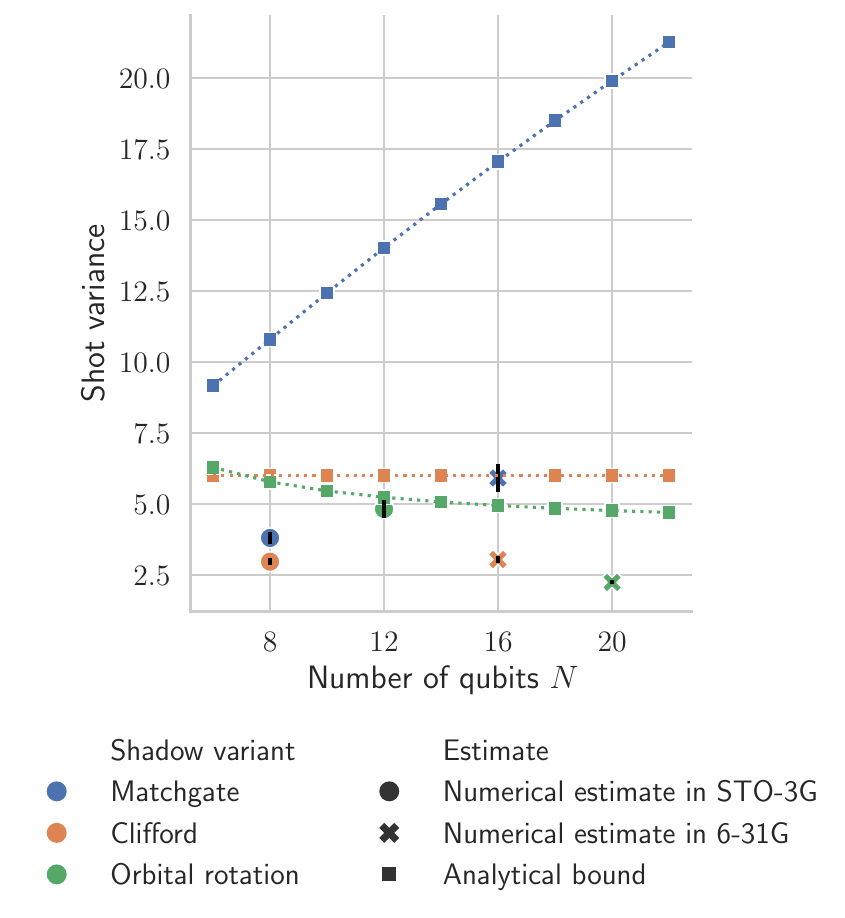}
    \caption{Numerical estimation of the average shot variances of all overlaps with computational basis states of the matchgate, Clifford and orbital rotation shadows for approximate ground states of $\ce{H4}$ in STO-3G basis ($N=8$, $\eta=4$) and in 6-31G basis ($N=16$, $\eta=4$) for $10^4$ shots in comparison with the analytical bounds of the variances.  Offset in $N$ of orbital rotation shadow estimation is due to the reference state requiring $N+\eta$ qubits. Dotted lines serve as a guide to the eye.}
\label{fig:lowmatchgate_scaling_varince}
\end{figure}

We further calculate the average shot variance of the three flavors of shadow estimation schemes (Matchgate, Clifford, Orbital rotation) for all computational basis states with the correct particle number for an approximate ground state of $\ce{H4}$ in the STO-3G and 6-31G basis obtained with VQE. We compare the numerical outcome  against their analytical bounds of their respective estimators in \cref{fig:lowmatchgate_scaling_varince}. The bounds are given by $4 \, b(N,\eta=4)$ from \cref{eq:kiannabound} (matchgate), $3 \operatorname{Tr}(O^2)$ from \cref{eq:clifford_var} (Clifford) and $4 \,\mathbb{E}_{\vec{p}, \vec{q}}\left[\mathrm{Var}\left[\left\langle\hat{D}_{\vec{q}}^{\vec{p}}\right\rangle\right]\right]$ from \cref{eq:lowbound} (Orbital rotation). 
The factors of 4 for matchgate and orbital rotation shadows in the analytical bounds are due to a factor of 2 in the estimation of the overlap with Slater determinants from   \cref{eq:kiannafactor2} and \cref{eq:overlapfromrdmlow}. The bound given for the orbital rotation shadow is bounding the average shot variance of the entries of the $\eta$-RDM \cite{low2022classical}, while the bounds in Clifford and matchgate shadows apply to the variance of individual observables; for further investigation, see \cref{sec:maxvarianceshadows}. We investigate the scaling behavior for fixed $\eta$ and $k$. The matchgate shadows scale $\mathcal{O}(N^2)$ while both the Clifford and orbital rotation shadows stay bounded, the orbital rotation shadow estimation shot variance even decreases with increasing $N$. This comes at the cost of increasing the qubit requirement by $\eta$, an increase in circuit depth for the orbital rotation shadows, and a classically inefficient post-processing step for the Clifford shadows. For the smaller basis, STO-3G, the matchgate shadow attains a lower shot variance than the orbital rotation shadow, while for the larger 6-31G basis, the orbital rotation shadow achieves the lowest average shot variance of $\approx 2.23$, roughly a factor of 2 lower than the numerical bound.

Up to this point, we studied the error scaling of the various methods assuming we are interested in calculating single overlaps. However, in QC-AFQMC,  the same shadows are used to estimate the overlaps between the trial wavefunction and walker states over many imaginary time steps. Although the walker states fluctuate throughout the AFQMC run, they maintain a significant overlap and cannot be considered independent. This raises the question of whether covariances might have an impact when the same classical shadows are reused to calculate overlaps between the (dependent) walker states and the VQE output.

To answer this question, we gather 100 independent shadows, each generated by averaging $100$ snapshots. We  use each shadow to compute the overlap between the VQE output state and (i) all computational states and (ii) 1000 walker states obtained from propagating a single walker 1000 imaginary time steps. Subsequently, we use the data to determine the covariance matrices for both sets of states. In \cref{fig:covariances}, we present our findings for the Clifford shadows, the matchgate shadows, and the orbital rotation shadows. Our findings reveal that for all ensembles, no covariances are present when estimating overlaps of computational states (right); however, covariances play a significant role when estimating overlaps with respect to walker states (left). As walker states have significant overlap with each other, see \cref{fig:walker_overlaps}, this observation is not unexpected but has to be taken into account when running the algorithm.
However,  we note that many techniques exist to reduce covariances from statistical estimates, such as resampling techniques \cite{efron1979bootstrap} or the use of Cholesky transformations.  

\section{QC-AFQMC under a finite number of measurements}\label{sec:qc_afqmc_finite}
In the following, we evaluate the performance of QC-AFQMC under the assumption that only a finite number of measurements can be taken on the quantum computer. 

\subsection{Propagation of errors}

The task of simulating the estimation of overlap between a VQE output state and walker states via classical shadows  becomes computationally demanding for larger systems due to the many overlaps required in the propagation, \cref{eqn_phase_diff_phaseless}, the calculation of the local energy, \cref{eqn_local_energy}, and calculation of the force bias, \cref{eqn_force_bias}. Furthermore, modern and efficient AFQMC implementations do not evaluate the local energy as in \cref{eqn_local_energy}, but rather leverage more efficient representations via Green's functions \cite{malone2022ipie}. To emulate the impact of overlap errors (and finite measurements) on the performance of AFQMC when applied to larger systems, we introduce an error model. This  allows us to carry out efficient AFQMC calculations in the presence of finite measurements.
As all measurement methods generate estimates, which follow a Gaussian error model, we draw a random error with mean $0$ and standard deviation $\epsilon$ and add it to the exact overlap.
Using Gaussian errors in combination with standard error propagation techniques, we propagate the error from the overlap calculation to the local energy calculation and the force bias calculation. In the following, we show this propagation for the local energy. For more details and the discussion of the error propagation to the force bias, we  refer to \cref{app:noise_model}.

As outlined in \cref{sec:overlaps_in_afqmc}, the computation of the local energy in \cref{eqn_local_energy} requires the estimation of up to $N^4$ overlaps on the quantum computer. Propagating the errors from all overlap estimations yields the following error in the local energy:
\begin{align}
    \epsilon_{\mathrm{LE}}^2 &\leq \frac{\epsilon^2\lambda^2}{\braket{\Psi_\mathrm{T}}{\phi}^2}+\epsilon^2 \matrixel{\Psi_\mathrm{T}}{H}{\phi}^2\leq\frac{2 \epsilon^2\lambda^2}{\braket{\Psi_\mathrm{T}}{\phi}^2}\,.
    \label{eq:local_energy_error}
\end{align}
Here, $\lambda^2=\left(\sum_{pq} |h_{pq}|^2 + \sum_{pqrs} |h_{pqrs}|^2\right)$  depends on the 2-norm of the Hamiltonian terms. We note that this error estimate decreases when restricting the sum to elements that correspond to reachable determinants within one- and two-body excitations from the walker state, see \cref{app:noise_model} for details.
In contrast, the error in variational quantum algorithms such as VQE, depends on the 1-norm of the Hamiltonian terms~\cite{tilly2022variational,bharti2022noisy,cerezo2021variational}. In molecular Hamiltonians where many terms are small, the here found 2-norm can be significantly smaller than the 1-norm.  
Assuming error-free walker weights and disregarding auto-correlation, we can propagate the error to the final estimate of the ground state energy, finding 
\begin{align}
    \epsilon_{\mathrm{E}}=\frac{\epsilon_{\mathrm{LE}}}{\sqrt{N_\mathrm{B}}\sqrt{N_{w}}}
    \label{eq:afqmc_error}\,,
\end{align}
where $N_\mathrm{B}$ and $N_{w}$ is the number of blocks and walkers respectively. Ignoring potential auto-correlations and other statistical issues, rather than allocating more measurements on the quantum computer, one could augment the number of blocks or walkers, albeit at the expense of escalating computational efforts on the classical computer.

\begin{figure}
    \centering
    \includegraphics[width=1\columnwidth]{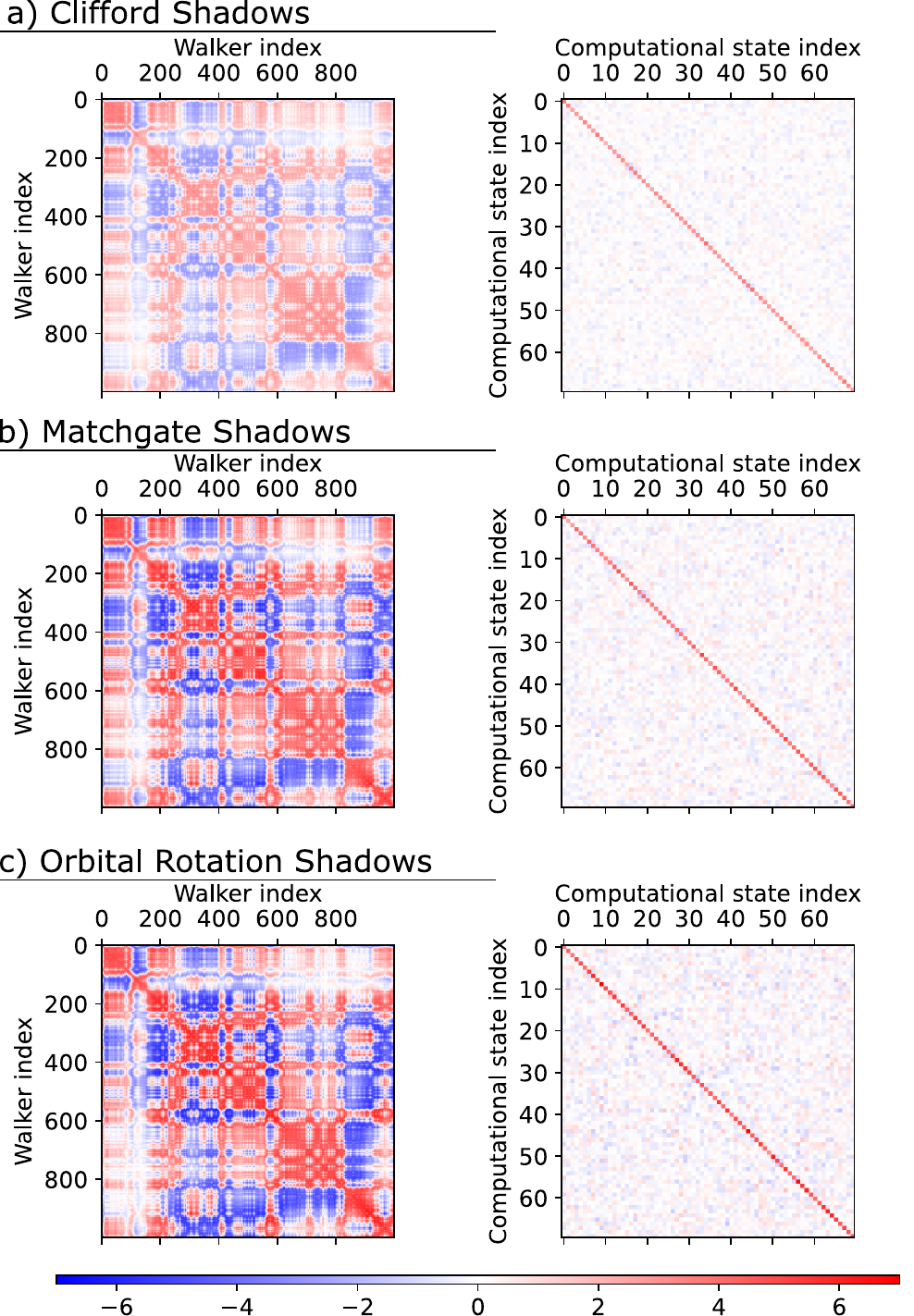}
    \caption{Single-shot covariances using 100 independent shadows estimates from the $\ce{H4}$ in STO-3G, each obtained by averaging $100$ snapshots. (left) The covariance matrix obtained when using the shadows to estimate the overlaps of 1000 walker states obtained from 1000 imaginary time steps. (right) The covariance matrix when using the same shadows to estimate the overlaps of the 70 computational basis states with correct particle number. The plots (a), (b), and (c) correspond to Clifford shadows, matchgate shadows, and orbital rotation shadows, respectively. }
    \label{fig:covariances}
\end{figure}

\subsection{Numerical simulations}

\begin{figure}
    \centering
    \includegraphics[width=1\columnwidth]{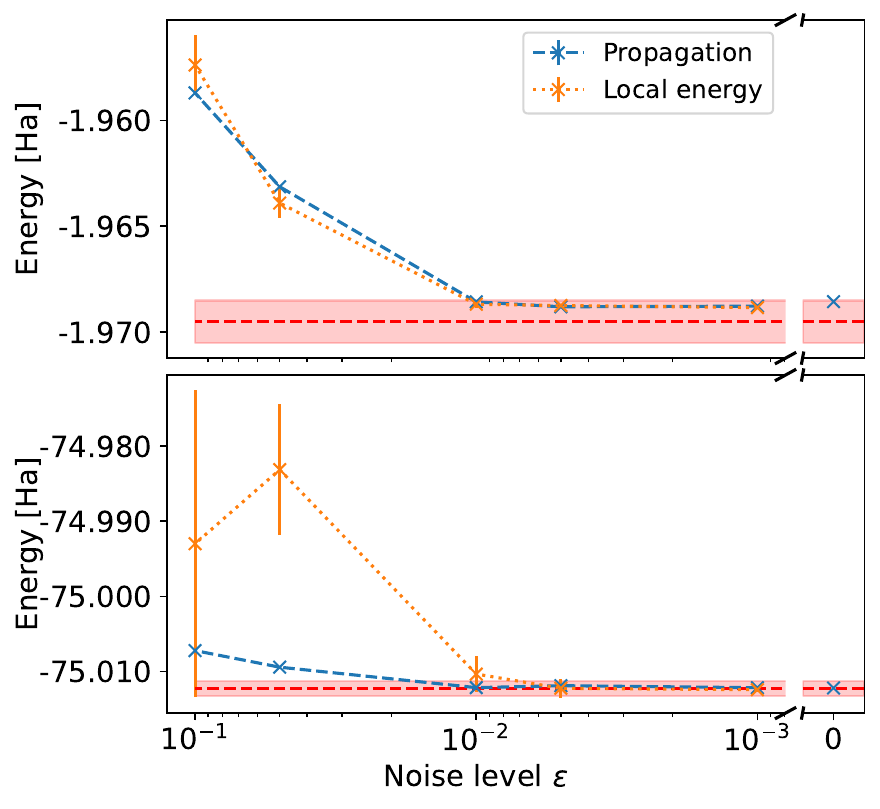}
    \caption{AFQMC energy estimates for two systems: (top) the $\ce{H4}$ square in STO-3G basis set, and (bottom) $\ce{H2O}$ in STO-3G basis set. We investigate the impact of varying noise levels $\epsilon$ in the overlap estimation. For comparison, we apply finite measurements in two scenarios: (1) only during propagation (blue curve), and (2) for estimating all quantities (orange curve). The dashed red curve represents the exact ground state energy for each system, while the red shaded area corresponds to energies which lie within chemical accuracy with respect to the ground state energy.}
    \label{fig:afqmc_estimates_h2_h2o}
\end{figure}
As a first test system, we employ the same system as in Sec.~\ref{sec:statistical_properties}; a $\ce{H4}$ square with a side length of $1.23$\AA{}, a frequently examined small system for correlated methods due to its present non-trivial correlations \cite{huggin2022unbiasing, amsler2023classical, baek2023say}. To simulate the VQE circuit on a classical computer, we use a minimal basis set, STO-3G. We implement the previously described QNP-VQE with $4$ layers. Upon optimizing the parameters using L-BFGS-B \cite{liu1989limited}, we find a VQE energy of $-1.962 \mathrm{Ha}$, which is $7 \mathrm{mHa}$ away from the exact ground state energy. It is worth mentioning that achieving a VQE value closer to the exact ground state energy for this small system is possible by increasing the number of layers and allowing more optimization cycles. However, our primary objective is to examine the impact of erroneous overlaps on the performance of QC-AFQMC rather than exploring the convergence of the VQE ansatz or comparing the performance of QC-AFQMC to classical quantum chemistry methods. 

Following this, we utilize the VQE output state as the trial wavefunction for AFQMC calculations. For all AFQMC calculations, we use the Python package ipie~\cite{malone2022ipie} with $N_{w}=1000$ walkers, $N_\mathrm{B}=1000$ blocks with $N_{\Delta\tau}=20$ imaginary time steps per block with an imaginary time step size of $\Delta\tau=0.005$. To estimate the AFQMC energy, we disregard the first $200$ blocks as the equilibration period.  \cref{fig:afqmc_estimates_h2_h2o}~(top) displays the AFQMC estimates in dependence of the noise level $\epsilon$ used. As illustrated in the plot, introducing errors into the propagation, \cref{eq:ovl}, results in a higher energy estimate than the error-less AFQMC calculation.  When incorporating errors into the propagation, \cref{eq:ovl}, the calculation of the force bias, \cref{eqn_force_bias}, and the estimation of the local energy, \cref{eqn_local_energy}, we observe a similar pattern as before; however, the errors in the estimate of the AFQMC energy are significantly larger than previously. 
It is worth mentioning that we did not employ reblocking methods to eliminate autocorrelation when calculating the AFQMC energy estimates. Statistical errors on the overlap propagate into the energy estimates at each step, causing fluctuations and making detecting autocorrelations more challenging. To ensure a fair comparison across all runs, we disregarded the effect of any autocorrelations.

As the second, slightly-larger test system, we use $\ce{H2O}$ with the geometry from the HEAT dataset \cite{tajti2004heat}, $r_\mathrm{OH}=0.95623$\AA{} and $\theta_\mathrm{HOH}=104.25^\circ$. We again exploit a minimal basis, STO-3G, and use the QNP-VQE ansatz with 12 layers to generate a trial wavefunction. After optimizing the parameters, we find an energy of $-75.002$ Ha, which is $10$ mHa away from the exact ground state energy. As before, we use the VQE output as a trial wavefunction in AFQMC, while exploring varying noise levels to estimate the overlaps. We report the results in \cref{fig:afqmc_estimates_h2_h2o}~(bottom). When reaching a noise level of $10^{-2}$, the AFQMC algorithm converges to an energy within chemical accuracy. However, the larger errors in the local energy estimation make the final AFQMC energy less predictable.

To study the effect of finite measurements on larger systems, we examine the $\ce{H4}$  in larger basis, cc-pVDZ and cc-pVTZ. As this maps to $40$ and $128$ spin orbitals and correspondingly the same number of qubits, running VQE  is not feasible anymore. Instead, we use PySCF's \cite{PySCF:2018} heat-bath CI implementation to generate trial wavefunctions. For both basis sets, we focus on finding the lowest singlet state. Subsequently, we use the trial wavefunction inside AFQMC with varying noise levels $\epsilon$.
We summarize our results in \cref{tab:afqmc_results}. For both basis sets, we find that the AFQMC energy estimates converge to the wrong value when the overlap errors used in the propagation increase. When incorporating overlap errors in the local energy estimation as well, we find that the errors increase faster when using the larger basis set due to \cref{eq:local_energy_error}.

\begin{table}[t!]
\centering
\begin{tabular}{c|cc}
\hline
 \multicolumn{3}{c}{AFQMC energy estimates of the \ce{H4} square (cc-pVDZ)}\\
\hline
\hline
\multicolumn{1}{c|}{Trial energy}       & \multicolumn{2}{c}{-2.00476}                         \\  \hline
\multicolumn{1}{c|}{Overlap error $\epsilon$} & \multicolumn{1}{c|}{Propagation} & \multicolumn{1}{c}{Full} \\ \hline
\multicolumn{1}{c|}{0.1}         & \multicolumn{1}{c|}{-2.01150(1)} & -2.01354(346)                 \\ \hline
\multicolumn{1}{c|}{0.01}        & \multicolumn{1}{c|}{-2.01207(2)} & -2.01102(94)                        \\ \hline
\multicolumn{1}{c|}{0.001}       & \multicolumn{1}{c|}{-2.01239(6)} & -2.01270(24)                            \\ \hline
\multicolumn{1}{c|}{No error}       & \multicolumn{1}{c|}{-2.01245(4)} & -2.01245(4)                          \\ 
\end{tabular}
\newline
\vspace*{5mm}
\newline
\begin{tabular}{c|cc}
\hline
 \multicolumn{3}{c}{AFQMC energy estimates of the \ce{H4} square (cc-pVTZ)}                         \\  
\hline
\hline
\multicolumn{1}{c|}{Trial energy}       & \multicolumn{2}{c}{-2.01786}\\
 \hline
\multicolumn{1}{c|}{Overlap error $\epsilon$} & \multicolumn{1}{c|}{Propagation} & \multicolumn{1}{c}{Full} \\ \hline
\multicolumn{1}{c|}{0.1}         & \multicolumn{1}{c|}{-2.02726(1)} & -2.01055(1118)                 \\ \hline
\multicolumn{1}{c|}{0.01}        & \multicolumn{1}{c|}{-2.02922(1)} & -2.02434(368)                        \\ \hline
\multicolumn{1}{c|}{0.001}       & \multicolumn{1}{c|}{-2.03275(1)} & -2.03190(182)                            \\ \hline
\multicolumn{1}{c|}{No error}       & \multicolumn{1}{c|}{-2.03399(7)} & -2.03399(7)                          \\ 
\end{tabular}
\caption{Energy estimates derived from AFQMC calculations of $\ce{H4}$ using the cc-pVDZ (top) and cc-pVTZ (bottom) basis sets, considering different overlap errors ($\epsilon$). The \textit{Propagation} column presents estimates from AFQMC runs with errors affecting only the overlaps in the importance function, \cref{eq:ovl}. Conversely, the \textit{Full}  column displays results where all overlap calculations are influenced by errors. All energies are reported in Hartree.}
\label{tab:afqmc_results}
\end{table}

\section{Discussion about classical \& quantum costs}\label{sec:costs}

As mentioned in \cref{sec:overlaps_in_afqmc}, the overlaps for the importance function need to be calculated $N_{w}N_\mathrm{B}N_\mathrm{\Delta\tau}\sim \mathcal{O}(1)$ many times in the AFQMC run. There may be some implicit dependence on system size with regard to how many walkers and how long of an evolution are required, but we ignore this for simplicity. The force bias requires $\mathcal O(N^2)$ overlaps for each walker to be calculated; thus, it needs $N_{w}N_\mathrm{B}N_\mathrm{\Delta\tau}\mathcal O(N^2)\sim \mathcal{O}(N^2)$ calculations of the overlap. Finally, the local energy estimate scales $\mathcal{O}(N^3)$ (when using compression methods) but must only be calculated $N_\mathrm{B}$-many times. These are the base factors for the scaling of calculating each quantity regardless of the method. We first consider the classical cost, summarized in  \cref{table:classical_scaling}, and then the quantum cost, summarized in  \cref{table:quantum_scaling}. 
\begin{table}[tb]
\begin{tabular}{c|c|c|c}
\hline
 \multicolumn{4}{c}{Scaling of post-processing cost} \\
\hline\hline
Method    & Single Overlap            & Local energy       & Force bias         \\ \hline
Hadamard  & $1$   & $N^3$ & $N^2$ \\ \hline
Clifford  & $2^N$ & $2^N$ & $2^N$ \\ \hline
Matchgate & $N^4$ & $N^7$ & $N^6$ \\ \hline
Orbital rot.  & $N^5$ & $N^{8}$ & $N^7$ \\ 
\end{tabular}
\caption{The scaling of the classical post-processing cost to estimate the overlap, the force bias for, and the local energy of a single walker. Only showing the largest scaling factor.}
\label{table:classical_scaling}
\end{table}

On top of the previously noted quantities of how many overlaps needed, we focus now on the number of classical operations required to calculate one overlap. In the case of the Hadamard test, the overlap is calculated by a simple average of samples and is not dependent on the system size. Thus, this method scales with $\mathcal{O}(1)$ and thus is independent of the system size.

In the case of Clifford shadows, the walkers, which are arbitrary Slater determinants, are described by an exponentially scaling linear combination of computational basis states, $\ket{\phi}=\sum_\alpha c_\alpha \ket{b_\alpha}$; thus, calculating the overlap scales exponentially with system size. Further, this needs to be performed for each snapshot. Thus, the scaling is $\mathcal{O}(2^N)$.

In the case of matchgate shadows, the calculation is dominated by the coefficient estimation performed by polynomial interpolation, which  scales  as $\mathcal{O}(N^4)$ because interpolation requires $\mathcal{O}(N)$-many calculations of a Pfaffian, of which the calculation scales as $\mathcal O(N^3)$ \cite{wimmer2012algorithm}. In orbital rotation shadows, this is extended to a grid of $N^2$ points for two-dimensional polynomial interpolation, therefore scaling as $\mathcal{O}(N^5)$. We further note that the default precision is typically insufficient as the system size grows for interpolating the polynomial coefficients. One needs to increase the numerical precision to get a sufficiently accurate estimate of the coefficients, which we discuss in more detail in \cref{sec:numerical_considerations}. 
Keeping the error energy estimates in AFQMC constant when increasing the system size typically requires $\mathcal{O}(N^2)$ samples of energy estimates \cite{lee2022twenty}. For a fixed number of walkers $N_{w}$ and imaginary time steps $N_{\Delta\tau}$, the classical computational cost to post-process all overlaps scales like $\mathcal{O}(N^{9})$ when using matchgate and $\mathcal{O}(N^{10})$ when using orbital rotation shadows. 
We also note that it might be beneficial to determine the overlaps of all computational states with the right symmetries and classically calculate the overlaps with the walker states. While this approach would technically scale exponentially with the system size, for certain regimes (e.g., skinny spaces), it might require less computational cost than determining the walker overlaps one by one.

\begin{table}[t!]
\begin{tabular}{c|c|c|c}
\hline
 \multicolumn{4}{c}{Scaling of the number of measurements in the}  \\\multicolumn{4}{c}{overlap estimation} \\
\hline
\hline
Method & Overlap & Local energy  & Force bias              \\ \hline
Hadamard   & $N_{w}N_\mathrm{B}N_{\Delta\tau}/\epsilon^2$ & $N^3N_{w}N_\mathrm{B}/\epsilon^2$ & $N^2N_{w}N_\mathrm{B}N_{\Delta\tau}/\epsilon^2$ \\ \hline
Clifford   & $1/\epsilon^2$ & $1/\epsilon^2$ & $1/\epsilon^2$ \\ \hline
Matchgate  & $\sqrt{N}/\epsilon^2$ & $\sqrt{N}/\epsilon^2$ & $\sqrt{N}/\epsilon^2$ \\ \hline
Orbital rot.  & $1/\epsilon^2$ & $1/\epsilon^2$ & $1/\epsilon^2$ \\ 
\end{tabular}
\caption{The $\tilde{\mathcal O}$-scaling (ignoring $\log$ terms) of the number of measurements to estimate the various objects in QC-AFQMC in scaling. Here, $\epsilon$ represents the additive error of each of the overlaps.}
\label{table:quantum_scaling}
\end{table}

The quantum computational cost of each method has two considerations: the number of copies of the trial state that are needed and the additional depth of the circuit required to perform the discussed quantum measurement protocols. We first look at the number of copies of the state required to calculate the overlaps for all walkers in the system, summarized in \cref{table:quantum_scaling}. Then, we discuss the circuits required for each case.

When using the Hadamard test, the number of samples needed is determined by the desired accuracy of the result $\mathcal{O}(1/\epsilon^2)$; however, one needs new copies of the trial state for every overlap calculation, in other words for every walker state, $N_{w}N_\mathrm{B}N_{\Delta\tau}$ in the total algorithm for the calculation of the importance sampling. In the case of calculating the local energy, $\mathcal{O}(N^3)$ copies per walker are required. Each term in the force bias also requires multiple copies of the trial state, which needs $\mathcal{O}(N^2)$ many overlap calculations per walker. In the case of Clifford shadows, the number of samples required scales with $\mathcal{O}(\log (N_{w})/\epsilon^2)$. In the case of matchgate shadows, the number of samples grows according to $\mathcal{O}(\sqrt{N}\log(N)\log (N_{w})/\epsilon^2)$. In both cases, due to the covariance problem of the walker states throughout the evolution, additional copies may be needed, but most likely only to a constant prefactor, to mitigate the effects of the covariance problem. We note that $\epsilon$ corresponds to the error made in the overlap estimation. As shown in Eq.~(\ref{eq:afqmc_error}), to bound the error in the AFQMC energy, the error has to decrease according to the 2-norm of the Hamiltonian terms.

The scaling of the circuit depth also depends on the method used to calculate the overlaps. As seen in \cref{fig:hadamard_circuit}, when using the modified Hadamard test, the circuit requires controlled versions of the state preparation circuit of the two states. Since the scaling of the controlled unitary acting on the full system scales with a high polynomial, one could employ methods that can replace the controlled unitary with an uncontrolled unitary \cite{huggins2020non,lu2021algorithms,russo2021evaluating}. In our case, the unitaries prepare the walker state and the trial wavefunction. The walker state is an arbitrary Slater determinant, which can be efficiently prepared with the latest algorithm using $\mathcal O (\eta\log^2(N))$ gate depth \cite{wecker2015solving,kivlichan2018quantum,google2020hartree,chee2023shallow}. The ansatz used for the trial state can be chosen to be efficient; however, the circuits for the walker states are still beyond the hardware available in the NISQ era. In the shadow technique framework, the depth of the circuit is dependent on the ensemble being sampled. In the case of $N$-qubit Clifford circuits, the Clifford unitaries can be implemented in $\mathcal O(N\log (N))$ depth using fully-connected topology \cite{berg2021simple}. In the case of matchgate circuits, the matchgate circuits that are also Clifford circuits can be used as an ensemble. Thus, the implementation of matchgate circuits is also efficient. Orbital rotation unitaries can be implemented in $\mathcal O(N)$ \cite{reck1993experimental}.

\section{Conclusion}

In this work, we studied the applicability of QC-AFQMC in terms of the number of measurements required from the quantum computer and the post-processing costs on the classical computer. We used various state-of-the-art measurement schemes to calculate the required overlaps between the non-orthogonal Slater determinants and the trial wavefunction prepared on the quantum computer and examined their statistical behavior. Numerical simulations revealed that unwanted covariances are present between estimates of walker overlaps at different imaginary time steps. Moreover, we exploited an error model based on Gaussian errors in the overlap calculations to analyze the error propagation to the AFQMC energy estimates. We found that the error of the AFQMC energy estimate can be upper bounded by the 2-norm of the Hamiltonian terms. This is in contrast with variational quantum algorithms which typically depend on the 1-norm.
We further discussed the classical post-processing cost of the various measurement methods and found that the post-processing of all required overlaps in QC-AFQMC scales with $\mathcal{O}(N^9)$ using matchgate shadows and $\mathcal{O}(N^{10})$ using orbital rotation shadows. This scaling of the current version of the algorithm hinders future applications in industrial settings.

Several steps are required to make QC-AFQMC viable for industrial applications. Firstly, regarding quantum computing, the development of quantum algorithms to generate classically intractable trial wavefunctions is vital for the future success of the method. 
Secondly, it is crucial to decrease the post-processing cost associated with the method substantially. Moreover, a comparison where the additional classical cost is spent on creating trial classical wavefunctions, e.g., from selected CI methods or low-bond-dimension DMRG calculations, could shed light on the application range of QC-AFQMC.
Next to advancements in the algorithm, most importantly, it is necessary to pinpoint industrial-relevant molecular systems where the method's high polynomial scaling can be justified in comparison to existing classical methods.

\section*{Acknowledgements}
We thank Ryan Babbush, Matthias Degroote, Florian Dommert, Bill Huggins, Michael Kaicher,  Fionn Malone, Clemens Utschig-Utschig,  and Christopher Wever for their valuable feedback on this manuscript. Additionally, we appreciate the members of the QUTAC Material Science working group for insightful discussions.
\bibliography{refs}

\newpage
\onecolumngrid
\appendix
\section{Error propagation}
\label{app:noise_model}
In the following, we show how an iid (independent and identically distributed) Gaussian error with mean $\mu=0$ and standard deviation $\epsilon$ in estimating the overlaps error propagates into AFQMC energy estimates and the force bias.

\subsection{Error in the energy estimation}

As shown in the SI of \cite{huggin2022unbiasing}, if one can measure the overlap between a trial wavefunction and a Slater determinant, $\braket{\Psi_\mathrm{T}}{\phi}$, the estimation of the numerator of the local energy is straightforward by exploiting the fact that the Hamiltonian only consists out of one and two-electron contributions
\begin{align}\label{eqn_local_energy_num}
    \matrixel{\Psi_\mathrm{T}}{H}{\phi}= &\sum_{pq}\braket{\Psi_\mathrm{T}}{\phi_p^q}\matrixel{\phi_p^q}{H}{\phi} + \sum_{pqrs}\braket{\Psi_\mathrm{T}}{\phi_{pq}^{rs}}\matrixel{\phi_{pq}^{rs}}{H}{\phi}\,,
\end{align}
where $\braket{\Psi_\mathrm{T}}{\phi_{p}^{q}}$ and $\braket{\Psi_\mathrm{T}}{\phi_{pq}^{rs}}$ are one- and two-body excitations from the walker state.
To bound the error of this, we first rotate into the orbital basis of the walker $\ket{\phi}$. This yields
\begin{align}
    \matrixel{\Psi_\mathrm{T}}{H}{\phi} = \sum_{pq} \braket*{\Tilde{\Psi}_\mathrm{T}}{\phi_{p}^{q}} \Tilde{h}_{pq}+\sum_{pqrs}\braket*{\Tilde{\Psi}_\mathrm{T}}{\phi_{pq}^{rs}} \Tilde{h}_{pqrs}\,,
\end{align}
where $\ket*{\Tilde{\Psi}_T}$, $\Tilde{h}_{pq}$ and $\Tilde{h}_{pqrs}$ are now given in the orbital basis of the walker.
Thus, we have to measure the overlaps between the (rotated) trial wavefunction and states obtained from one-body  and two-body excitations from the walker states,  $\braket{\Psi_\mathrm{T}}{\phi_{p}^{q}}$ and $\braket{\Psi_\mathrm{T}}{\phi_{pq}^{rs}}$ respectively.
As before, we assume that we can measure $\braket{\Psi_\mathrm{T}}{\phi}$ up to an additive error $\epsilon$. We moreover assume that covariances between estimates of overlaps of excited walker states can be neglected, as numerically shown in Fig.~\ref{fig:covariances}(b). Using propagation of errors, we find the error in the nominator of the local energy estimation. 
\begin{align}
    \epsilon_\mathrm{num.}<\epsilon \sqrt{\left(\sum_{pq} |\Tilde{h}_{pq}|^2 + \sum_{pqrs} |\Tilde{h}_{pqrs}|^2\right)}=:\epsilon\lambda\,.
\end{align}
As the 2-norm is invariant under unitary transformations, we can write $\lambda^2=\sum_{pq} |h_{pq}|^2 + \sum_{pqrs} |h_{pqrs}|^2$.
We note that this is an upper bound to the error, which we found by summing over all possible Hamiltonian coefficients. However, a given walker only connects via the Hamiltonian to a subset of all possible determinants, and thus, not all coefficients will appear in the sum. For example, if orbital $X$ is unoccupied in walker $\phi$, $\Tilde{h}_{pqrX}$ will not appear in the sum. This could be taken into account when limiting the indices in the sum to the feasible particle-hole excitation and particle-particle-hole-hole combinations. For cases where the number of electrons is small in comparison to the number of orbitals, this could provide a substantial improvement.

The denominator of the local energy is an overlap itself. Assuming again that we can estimate it up to the same additive error $\epsilon$ allows us to calculate the error on the local energy:
\begin{align}
    \epsilon_\mathrm{LE} &= E_\mathrm{loc}(\phi)\epsilon\sqrt{\frac{\lambda^2}{\matrixel{\Psi_\mathrm{T}}{H}{\phi}^2}+\frac{1}{\braket{\Psi_\mathrm{T}}{\phi}^2}}\\
    \Rightarrow \epsilon_{\mathrm{LE}}^2 &= \frac{\epsilon^2\lambda^2}{\braket{\Psi_\mathrm{T}}{\phi}^2}+\epsilon^2 \matrixel{\Psi_\mathrm{T}}{H}{\phi}^2\\
    &\leq \epsilon^2\lambda^2 \left(1+\frac{1}{\braket{\Psi_\mathrm{T}}{\phi}^2}\right)\leq\frac{2 \epsilon^2\lambda^2}{\braket{\Psi_\mathrm{T}}{\phi}^2}\,.
\end{align}
The error in estimating the local energy depends on the 2-norm $\lambda$ of the Hamiltonian and the overlap between the trial and the walker state $\braket{\Psi_\mathrm{T}}{\phi}$. When the overlap between the walker and the trial is exponentially decreasing, an exponential number of measurements will be required to estimate the local energy to constant error. 
In AFQMC, the local energies are used to generate an estimate of the total energy. 
\begin{align}
    E_{tot}=\frac{\sum w_\phi E_{\mathrm{loc}}(\phi)}{\sum w_\phi}
\end{align} 
Assuming error-less weights, the error of the total energy is only dependent on the error of the local energy and the number of walkers $N_{w}$ used,
\begin{align}
    \epsilon_{\mathrm{tot}}= \sqrt{\sum w_\phi^2 \epsilon_\mathrm{LE}^2 }\approx \frac{\epsilon_\mathrm{LE}}{\sqrt{N_{w}}}\,.
\end{align}

The final estimate of the AFQMC energy is given by averaging the total energy over many time steps (after equilibration). Ignoring reblocking techniques to reduce the autocorrelation between time steps, the final estimate can be expressed as
\begin{align}
    E_\textrm{AFQMC}=\frac{1}{N_\mathrm{B}}\sum_i^b E_\mathrm{tot}^{(i)}\,.
\end{align}
Using the error on the total energies, we find the error on the final AFQMC energy to be
\begin{align}
    \epsilon_\textrm{E}=\frac{\epsilon_{tot}}{\sqrt{N_\mathrm{B}}}=\frac{\epsilon_{LE}}{\sqrt{N_\mathrm{B}}\sqrt{N_{w}}}
\end{align}
Note, that this is only the error due to the finite measurements during the local energy estimation. The total error of the AFQMC sill includes the statistical Monte Carlo errors. Moreover, due to finite measurements in the propagation, the energy could converge to a wrong value. Moreover, as we discussed in \cref{subsec:classical_shadows}, covariances exist, and the errors cannot necessarily be assumed to be iid.

\subsection{Force bias}

Similarly to the local energy calculation, we can also propagate the error of the overlap calculations into the estimation of the force bias. As given in \cref{eqn_force_bias}, the force bias is defined as
\begin{align}
    \Bar{x}_\gamma=-\sqrt{\Delta\tau}\frac{\matrixel{\Psi_\mathrm{T}}{v_\gamma}{\phi}}{\braket{\Psi_\mathrm{T}}{\phi}}\,,
\end{align}
where $v_\gamma=i \sum_{pq}L^{\gamma}_{pq} a_p^\dagger a_q$ are one-body terms with $L^{\gamma}_{pq}$ being the Cholesky vectors. As before, this can be rewritten as
\begin{align}
    \Bar{x}_\gamma=-i\sqrt{\Delta\tau}\frac{\sum_{pq}\braket{\Psi_\mathrm{T}}{\phi_{p}^{q}}\matrixel{\phi_{p}^{q}}{L^{\gamma}_{pq}}{\phi}}{\braket{\Psi_\mathrm{T}}{\phi}}\,,
\end{align}
where $\ket{\phi_{p}^{q}}$ denotes a walker state with an excitation from the $p$ to the $q$ orbital. 
Assuming again an additive error $\epsilon$ for each of the overlap estimations, we find the error of the local energy to satisfy
\begin{align}
    \epsilon_{\Bar{x}_\gamma}=\Bar{x}_\gamma \epsilon \sqrt{\frac{\sum_{pq}|L^{\gamma}_{pq}|^2}{(\sum_{pq}\braket{\Psi_\mathrm{T}}{\phi_{p}^{q}}\matrixel{\phi_{p}^{q}}{L^{\gamma}_{pq}}{\phi})^2}+\frac{1}{\braket{\Psi_\mathrm{T}}{\phi}^2}} \, .
\end{align}

\section{Numerical considerations}\label{sec:numerical_considerations}

Polynomial interpolation is used to determine the coefficients of the polynomial $q(z)$ from \cref{eq:matchgateinversechannel}. We summarize a common method and discuss the numerical considerations when practically implementing this method.

The goal is to estimate the $n$-degree polynomial
\begin{equation}
    p(x)=c_0+c_1x+\cdots+c_{n-1}x^{n-1}+c_nx^n~,
\end{equation}
where the coefficients $\{c_i\}_{i=0}^n$ are unknown. One can form a system of linear equations by evaluating the polynomial at the points $\{x_i\}_{i=0}^m$ obtaining the outputs $\{p(x_i)=y_i\}_{i=0}^m$ described by the matrix vector multiplication
\begin{equation}
    \begin{bmatrix}
        &1&x_0&\cdots&x_0^{n-1}&x_0^n\\
        &1&x_1&\cdots&x_1^{n-1}&x_1^n\\
        &\vdots&\vdots&\ddots&\vdots&\vdots\\
        &1&x_m&\cdots&x_m^{n-1}&x_m^n\\
    \end{bmatrix}
    \begin{bmatrix}
        c_0\\
        c_1\\
        \vdots\\
        c_{n-1}\\
        c_n\\
    \end{bmatrix}
    =\begin{bmatrix}
        y_0\\
        y_1\\
        \vdots\\
        y_m\\
    \end{bmatrix}~,
\end{equation}
where the matrix is known as the Vandermonde matrix, $V$. This matrix is always invertible when $m=n$. Thus, one can obtain the unique solution of the coefficients $\boldsymbol{c}=V^{-1}\boldsymbol{y}$. The best choice for the points $\{x_i\}_{i=0}^m$ is the Chebyshev nodes which are defined in the interval $[-1,1]$ as
\begin{equation}
    x_i = \cos\left(\frac{2i-1}{2m}\pi\right)~.
\end{equation}

When implementing this, one encounters issues of numerical stability due to the high condition number of the Vandermonde matrix. When using standard \texttt{python} scripts, we saw that once the polynomial reaches a degree of 20-25, the instability is so significant that the results are no longer sufficiently accurate to machine precision (assumed to be $10^{-13}$). Therefore, one needs to increase the precision of the calculations used. 

\begin{figure}
    \centering
    \includegraphics[width=0.75\columnwidth]{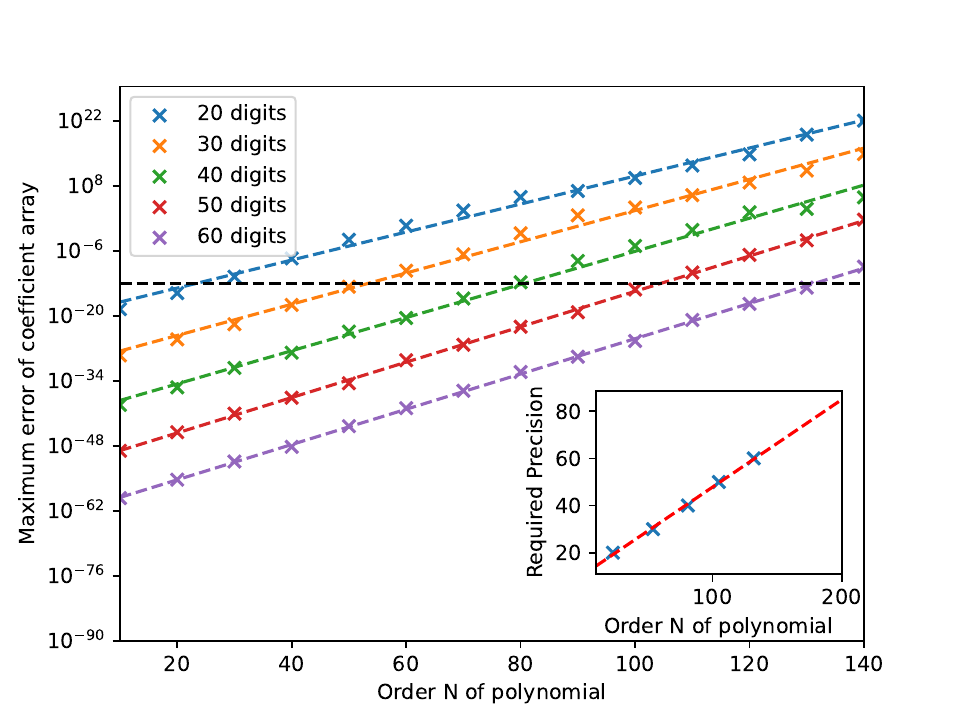}
    \caption{We calculated the maximum error of interpolating the polynomial coefficients as the degree of the polynomial increases for various digits of precision used in the calculations. The inset shows the required number of digits of precision to stay below machine error ($10^{-13}$) for various different degrees of polynomial and the line of best fit (the dashed red line).}
    \label{fig:numerical_stability}
\end{figure}

We used the package \texttt{mpmath}~\cite{mpmath} to increase the precision of the calculations. We calculated the maximum error of interpolating the polynomial coefficients as the degree of the polynomial increases for various digits of precision used in the calculations. As seen in \cref{fig:numerical_stability}, the error in the estimate for every level of precision grows as the order of the polynomial increases. As seen in the inset of \cref{fig:numerical_stability}, our analysis suggests that the required digits of precision grow linearly with the order of the polynomial.

\subsection{Covariances in overlap estimations}
\begin{figure}
    \centering
    \includegraphics[width=0.5\columnwidth]{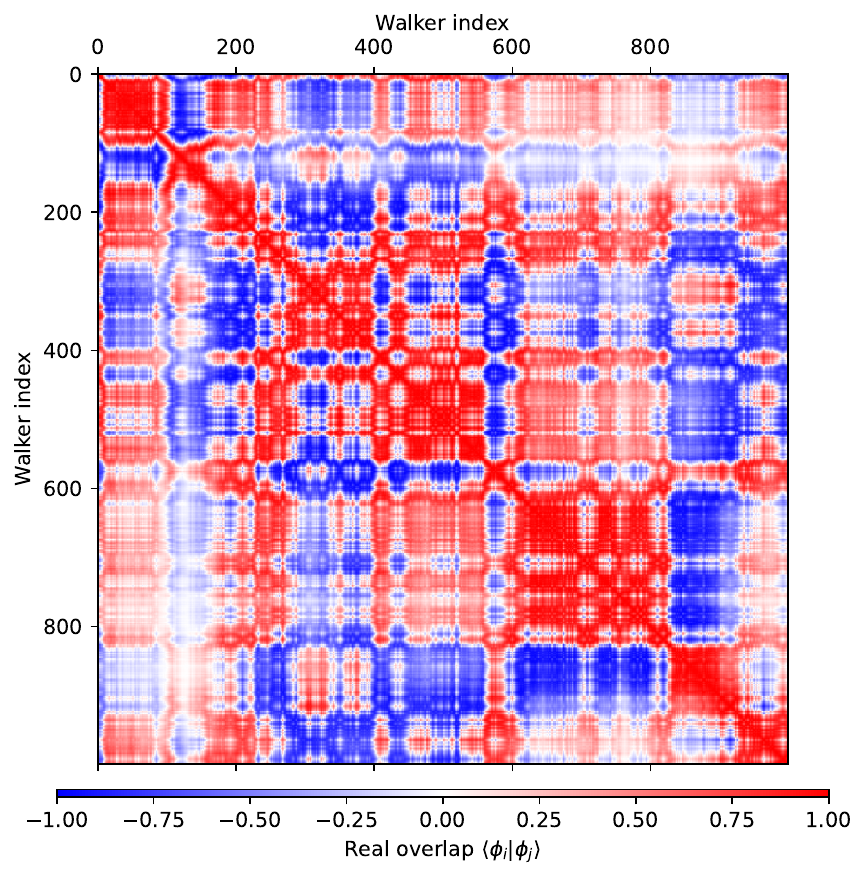}
    \caption{The overlap matrix $\braket{\phi_i}{\phi_j}$ between walkers at different time steps $i$ and $j$.}
    \label{fig:walker_overlaps}
\end{figure}

It is well known that for real-valued random variables $X_i,Y_j$ and constants $a_i,b_j$ the covariances between linear combinations of the random variables can be calculated as:
\begin{equation}
    \operatorname{Cov}\left(\sum_i a_i X_i, \sum_j b_j Y_j\right) = \sum_{ij} a_ib_j \operatorname{Cov}({X_i,Y_j})
\end{equation}
In our case, the overlap of a trial state and a general Slater determinant $\phi_a$ is a linear combination out of overlaps with the overlaps of computational basis states $c_i$, which have vanishing covariances as discussed above. For the covariance between estimates of two of these walkers $\phi_a,\phi_b$  we obtain 
    \begin{align}
            \operatorname{Cov}\left(\sum_i a_i c_i, \sum_j b_j c_j\right) &= \sum_{ij} a_ib_j \operatorname{Cov}({c_i,c_j}) \\
            &= \sum_{i} a_ib_i \operatorname{Cov}({c_i,c_i}) \\
            &= \sum_{i} a_ib_i \operatorname{Var}(c_i) \\
            &\approx \operatorname{Var}_\text{mean} * \sum_{i} a_ib_i \\
            &= \operatorname{Var}_\text{mean} \langle \phi_a | \phi_b \rangle\,,
    \end{align}
where we used that covariances between estimates of different computational basis states vanish and approximated all variances by the mean variance $\operatorname{Var}_\text{mean}$, which is a good approximation for a large number of shots as shown in \cite{scheurer23:_tailor_exter_correc_coupl_clust_quant_input}. We can see that the covariance of the overlap is approximately proportional to the overlap between the walkers $\langle \phi_a | \phi_b \rangle$, which can be seen when comparing the plot of the walker overlaps in Fig. \ref{fig:walker_overlaps} with the plot of the covariances in Fig. \ref{fig:covariances}.

\section{Bound of the variance of matchgate shadows}
The overlap of the trial wavefunction $|\Psi_\mathrm{T}\rangle$ with an arbitrary Slater determinant $|\phi\rangle$ can be estimated from \cite{wan2022matchgate}
\begin{equation}
\label{eq:kiannafactor2}
    \left\langle\Psi_\mathrm{T}|\phi\right\rangle=2 \operatorname{tr}\left(\left|\phi\right\rangle\langle {0}| \rho_\tau\right) ,
\end{equation}
where $\rho_\tau$ is the density matrix $|\tau\rangle \langle\tau|$ of the state in \cref{eq:tau_state}. The variance of $\operatorname{tr}\left(\left|\phi\right\rangle\langle {0}| \rho_\tau\right)$ in the matchgate shadow estimation scheme is then bounded by \cite{wan2022matchgate}
\begin{equation}
\label{eq:kiannabound}
    \left.\operatorname{Var}[\hat{o}]\right|_{O=|\phi\rangle\langle\mathbf{0}|} \leq b(N, \eta):=\frac{1}{2^{2 N}} \sum_{\substack{\ell_1, \ell_2, \ell_3 \geq 0 \\ \ell_1+\ell_2+\ell_3 \leq N}} \alpha_{\ell_1, \ell_2, \ell_3} \kappa\left(N, \eta, \ell_1, \ell_2, \ell_3\right),
\end{equation}
where $\alpha_{\ell_1, \ell_2, \ell_3}$ and $\kappa\left(N, \eta, \ell_1, \ell_2, \ell_3\right)$ are defined below, $N$ is the total number of fermionic modes and $\eta$ is the number of occupied fermionic modes.
\begin{equation}
\alpha_{\ell_1, \ell_2, \ell_3} :=
\frac{\left(\begin{array}{c}
N \\
\ell_1,\ell_2,\ell_3,N-\ell_1-\ell_2-\ell_3
\end{array}\right)}{\left(\begin{array}{c}
2N \\
2\ell_1,2\ell_2,2\ell_3,2\left(N-\ell_1-\ell_2-\ell_3\right)
\end{array}\right)}
    \frac{\left(\begin{array}{c}
2 N \\
2\left(\ell_1+\ell_3\right)
\end{array}\right)}{\left(\begin{array}{c}
N \\
\ell_1+\ell_3
\end{array}\right)} \frac{\left(\begin{array}{c}
2 N \\
2\left(\ell_2+\ell_3\right)
\end{array}\right)}{\left(\begin{array}{c}
N \\
\ell_2+\ell_3
\end{array}\right)}
\end{equation}
\begin{equation}
    \kappa\left(N, \eta, \ell_1, \ell_2, \ell_3\right):=2^\eta \sum_{j=0}^{\eta / 2}\left(\begin{array}{c}
\eta \\
2 j
\end{array}\right)\left(\begin{array}{c}
N-\eta \\
\ell_1-\eta / 2+j, \ell_2-\eta / 2+j, \ell_3-j, N-\ell_1-\ell_2-\ell_3-j
\end{array}\right)
\end{equation}

The factor of 2 in~\cref{eq:kiannafactor2} implies a factor of 4 in the variance when estimating overlap with Slater determinants in this scheme.
\section{Beyond average performance of shadow estimation}
\label{sec:maxvarianceshadows}
\begin{figure}
    \centering
\includegraphics[width=0.5\columnwidth]{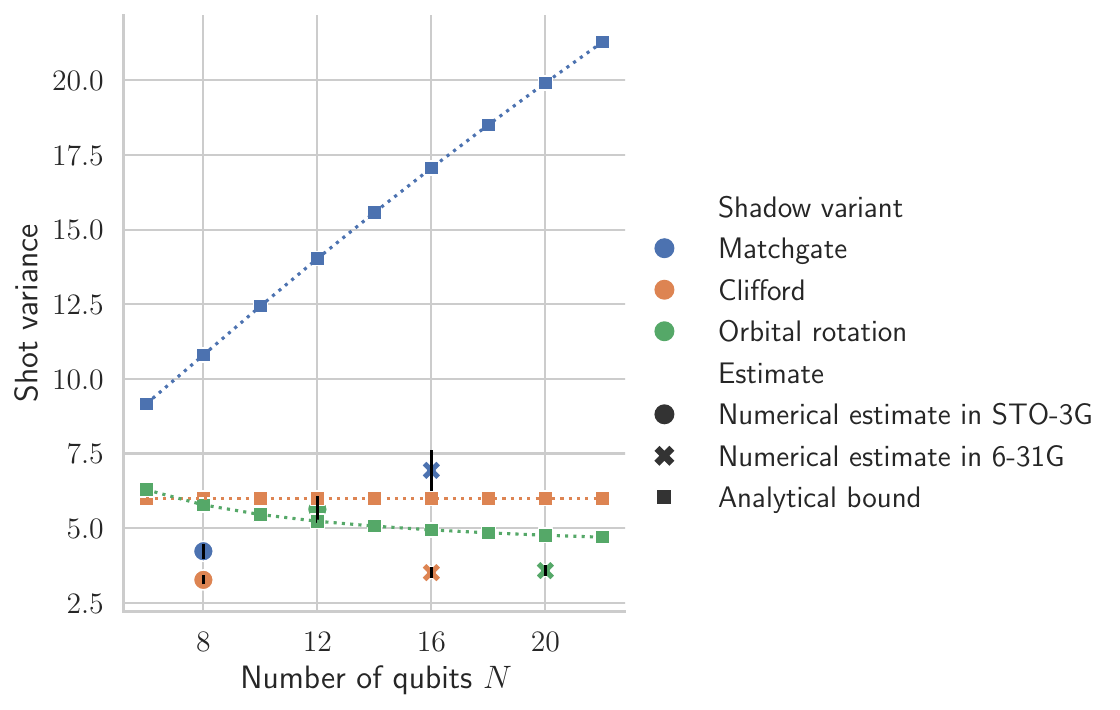}
\includegraphics[width=0.45\columnwidth]{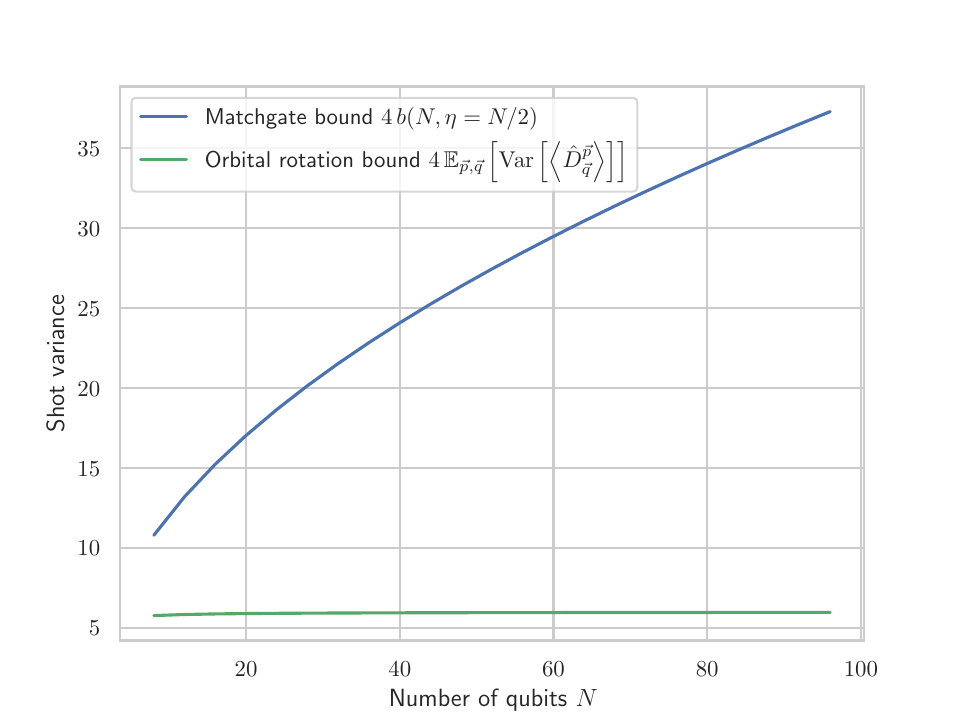}
    \caption{Left: Numerical estimation of the maximum shot variance of an individual computational basis state of the matchgate, Clifford and orbital rotation shadows for approximate ground states of $\ce{H4}$ in STO-3G basis ($N=8$, $k=4$, $\eta=4$) and in 6-31G basis ($N=16$, $k=4$, $\eta=4$) for $10^4$ shots in comparison with the analytical bounds of the variances. Dotted lines serve as a guide to the eye. Right: Scaling behavior of the shot variance bounds for N and $\eta=N/2$.}
\label{fig:max_variance}
\end{figure}

Numerical simulations are done via the Pennylane suite \cite{bergholm2022pennylane}, matchgate unitaries are implemented by a new developed efficient Majorana decomposition \cite{zhao2023grouptheoretic} and orbital rotation unitaries by a decomposition into Givens rotation \cite{reck1993experimental}. Post-processing schemes are implemented as described in \cite{wan2022matchgate,low2022classical}.
To investigate worst case performance of the estimation schemes, we plot the maximum variance of an individual computational basis state overlap in Fig. \ref{fig:max_variance} left of the same numerical simulation done in Fig. \ref{fig:lowmatchgate_scaling_varince}. We find that individual computational basis state overlaps can perform worse than the bound for the average variance for orbital rotation shadows in STO-3G. At higher $N$, even the individual elements do not violate the bound, and even the worst variance of an individual element performs better than the average variance of the matchgate shadows from Fig. \ref{fig:lowmatchgate_scaling_varince}. In Fig. \ref{fig:max_variance} right, we also plot the scaling behavior for the two bounds when $\eta$ is not fixed and allowed to scale with system size, here $\eta=N/2$. The bound of the orbital rotation shadow stays constant, while for increasing system size, the average shot variance of the matchgate shadow increases. Thus, all overlaps with Slater determinants can be extracted with a constant number of quantum measurements in the orbital rotation shadow.

\begin{figure}
    \centering  \includegraphics[width=0.5\linewidth]{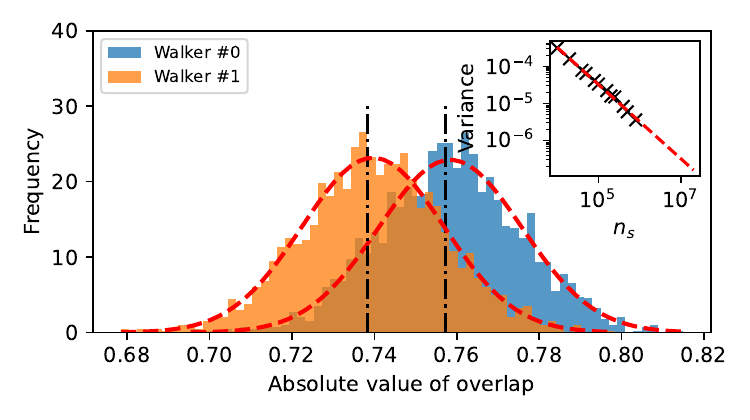}    \caption{Estimating overlaps with Clifford shadows. (main plot) The histograms show the distribution of the absolute values of the overlap estimates of two walker states w.r.t. a VQE output state ($\ce{H4}$) by conducting 2000 repeated shadow experiments, with each data point utilizing $10^4$ Clifford shadows. (inset) The standard deviation in the overlap distribution as the number of shadows increases.}
    \label{fig:clifford_scaling_varince}
\end{figure}
\end{document}